\documentclass[12pt]{iopart}

\usepackage{CJK}
\usepackage{graphicx} 
\usepackage{longtable} 
\usepackage{xcolor,soul} 
\usepackage{multirow} 
\usepackage{amsmath} 
\usepackage[hidelinks]{hyperref} 
\usepackage{bm} 
\newcommand{\red}[1]{\color{black}{#1} \color{black}}

\begin{document}
\begin{CJK*}{GBK}{ }

\title[]{How to identify and characterize strongly correlated topological semimetals}
\author{Diana M.\ Kirschbaum$^*$, Monika Lu\v{z}nik$^*$, Gwenvredig Le Roy, and Silke Paschen}
\address{Institute of Solid State Physics, Vienna University of Technology, Wiedner Hauptstr. 8-10, 1040 Vienna, Austria}
\end{CJK*}

\begin{abstract}
How strong correlations and topology interplay is a topic of great current interest. In this perspective paper, we focus on correlation-driven gapless phases. We take the time-reversal symmetric Weyl semimetal as an example because it is expected to have clear (albeit nonquantized) topological signatures in the Hall response and because the first strongly correlated representative, the noncentrosymmetric Weyl--Kondo semimetal Ce$_3$Bi$_4$Pd$_3$, has recently been discovered. We summarize its key characteristics and use them to construct a prototype Weyl--Kondo semimetal temperature-magnetic field phase diagram. This allows for a substantiated assessment of other Weyl--Kondo semimetal candidate materials.
We also put forward \red{scaling plots of the intrinsic Berry-curvature-induced Hall response vs the inverse Weyl velocity---a measure of correlation strength, and vs the inverse charge carrier concentration---a measure of the proximity of Weyl nodes to the Fermi level. They suggest that the topological Hall response is maximized by strong correlations and small carrier concentrations.} We hope that our work will guide the search for new Weyl--Kondo semimetals and correlated topological semimetals in general, and also trigger new theoretical work.
\end{abstract}

\section{Introduction}\label{intro}
Heavy fermion compounds are materials where itinerant and localized (typically $4f$ or $5f$) electrons coexist and, at low enough temperatures $T$, strongly interact via the Kondo effect. They are best known for the heavy effective masses of their charge carriers, the property that gave this class of materials its name \cite{Ste84.1,Hew97.1}. They are also known for their ready tunability. Small variations of an external (nonthermal) control parameter $\delta$ such as pressure or magnetic field lead to strong changes in the effective mass. Particularly drastic enhancements appear when approaching a quantum critical point where, at a critical value $\delta_{\rm c}$ of the control parameter, a second-order (typically antiferromagnetic) phase transition is just suppressed to $T=0$ \cite{Loe07.1}. The standard method to experimentally determine effective masses of heavy fermion metals is to measure a physical property at sufficiently low temperatures such that it exhibits Fermi liquid behavior. The effective mass can then be determined by comparison with the corresponding theoretical Fermi liquid expression, e.g., $C(T)=\gamma T$ for the electronic specific heat, $\Delta\rho(T)=AT^2$ for the electrical resistivity, or $\chi(T)=\chi_0$ for the magnetic susceptibility of the conduction electrons, where the Sommerfeld coefficient $\gamma$, the resistivity $A$ coefficient, and the Pauli susceptibility $\chi_0$ are all related to the effective mass \cite{Ste84.1,Hew97.1,Kad86.1,Jac09.1}. Upon approaching a quantum critical point, situated at $T=0$ and $\delta=\delta_{\rm c}$, these temperature dependences hold in ever narrower temperature ranges as they give way to non-Fermi liquid behavior emerging at the quantum critical point and extending in a fan-like shape into the $T(\delta)$ phase diagram \cite{Loe07.1,Sch99.1,Col01.1,Ste01.1,Kir20.1,Pas21.1}. 

It is important to note that the above relations hold for metals. Whereas most heavy fermion compounds are indeed metallic, there is a smaller subset of materials that display semiconducting properties. They are typically referred to as Kondo insulators \cite{All78.1,Tra84.1,Aep92.1,Buc94.1,Roz96.1,Tsu97.1,Ris00.1,Yam10.1,Si13.1}. In a simple mean-field picture, the insulating state arises due to the hybridization of the conduction electrons with the localized electrons, and the Fermi level lies within this hybridization gap. The periodic Anderson and Kondo lattice models are also known to exhibit such gaps \cite{Ono91.1,Pru00.1}; at half filling, where the lower hybridized band is fully occupied and the upper hybridized band is empty, a Kondo insulator results. The above Fermi liquid relations may still be meaningful if effects such as doping or off-stoichiometry move the Fermi level from within the gap into the conduction or valence band, or even into a conductive impurity band. In that case, the knowledge of the charge carrier concentration is needed to estimate the mass enhancement from experimental values of $\gamma$, $A$, or $\chi_0$. An alternative measure of correlation strength is the width of the gap (the narrower it is the stronger the correlations), but experimentally determined gap magnitudes have typically differed strongly depending on the quantity they were extracted from.

The field of Kondo insulators underwent a sudden revival as---with the advent of topological insulators \cite{Has10.1}---also topological Kondo insulators were proposed \cite{Dze10.1}. In this first work, a topologically nontrivial insulating state was found to result from the spin--orbit coupling associated with the
hybridization between the conduction and localized ($f$) electrons, in particular for certain positions of the renormalized $f$ level relative to the bottom of the conduction band and for certain crystal symmetries at the $f$ electron site. This proposal raised great interest and triggered massive efforts  \cite{Jia13.1,Neu13.1,Wen14.1,Kim14.2,Li14.1,Xu14.1,Tan15.1,Dze16.1,Nak16.1,Par16.1,Rac18.1,Pir20.1,Li20.2,Ais22.1}. Nevertheless, in spite of considerable progress, there is no broad consensus yet on the topological nature of the observed surface states. Part of the challenge derives from the fact that the surface of a Kondo insulator is a delicate object. The formation of the Kondo insulator gap requires the Kondo effect to operate, something that naturally fails on a surface, where the Kondo screening cloud is cut off. Secondly, the tools that have provided rapid progress in the field of noninteracting topological insulators, most notably angle-resolved photoemission spectroscopy (ARPES) in combination with density functional theory (DFT), are of limited use for Kondo insulators, both due to their narrow bandwidths and the absence of precise ab-initio methods. Finally, predictions for robust and readily testable experimental signatures of the expected topological surface states are scarce.

More recently, in a joint effort of experiment and theory, heavy fermion compounds with metallic topology have been advanced, at first the Weyl--Kondo semimetal \cite{Dzs17.1,Lai18.1,Gre20.1,Dzs21.1} and later the Weyl--Kondo nodal-line semimetal \cite{Che22.1}. They result from the interplay of the Kondo effect, strong spin--orbit coupling, and specific lattice symmetries, \red{and are strongly correlated analogs of the previously discovered noninteracting and weakly interacting Weyl semimetals \cite{Arm18.1, Has21.1}.} The Weyl--Kondo semimetal, which has Weyl point nodes, was theoretically demonstrated in a periodic Anderson model, with conduction electrons on a zincblende lattice, a simple noncentrosymmetric structure \cite{Lai18.1,Gre20.1}. The Weyl--Kondo nodal-line semimetal, by contrast, was found for conduction electrons on a 3D lattice of space group (SG) $Pmn2_1$ (No.\ 31) \cite{Che22.1}. For the considered commensurate filling, the nodes appear at the Fermi energy as the Kondo effect develops. The linear dispersion near the Weyl nodes is extremely flat, with the renormalized bandwidth given by the Kondo temperature. Experimentally, Weyl--Kondo semimetal behavior was first found in the heavy fermion compound Ce$_3$Bi$_4$Pd$_3$ \cite{Dzs17.1,Dzs21.1}, which crystallizes in a cubic, noncentrosymmetric and nonsymmorphic structure of SG $I\bar{4}3d$ (No.\ 220). Initial evidence for Weyl--Kondo nodal-line semimetal behavior was found in Ce$_2$Au$_3$In$_5$ \cite{Che22.1}, which forms in the orthorhombic, noncentrosymmetric, and nonsymmorphic structure of SG $Pmn2_1$ (No.\ 31). Ce$_3$Bi$_4$Pd$_3$ displays ``giant'' signatures of nontrivial topology, which was attributed to the effect of strong correlations due to the Kondo effect \cite{Dzs17.1,Dzs21.1}.

In this perspective, we will highlight these features to facilitate the identification of other representatives of this new class of materials. We will also examine the relationship between the size of the topological responses and the strength of electronic correlations, which can be used to verify experimental interpretations. The paper is organized as follows. We summarize the key features of Weyl--Kondo semimetal phase in Section \ref{char} and discuss various candidate materials in Section \ref{cand}. In Section \ref{quant} we explain how the correlation strength can be quantified in these materials and in Section \ref{topo} we investigate the relationship between correlation strength and the size of the topological responses. In Section \ref{outlook} we summarize and discuss our findings, and provide an outlook.

\section{Characteristics of the Weyl--Kondo semimetal}\label{char}
The Weyl--Kondo semimetal is a new state of matter put forward in a joint effort of experiment \cite{Dzs17.1,Dzs21.1} and theory \cite{Lai18.1,Gre20.1}. It may form in systems with preserved time reversal symmetry but broken inversion symmetry. As it is the currently best-established gapless topological state driven by strong electron correlations, it is the focus of this perspective paper. The understanding that results from the above works is that Weyl nodes, which are already present in the noninteracting bandstructure, become part of the Kondo resonance at low temperatures and thus appear in the immediate vicinity of the Fermi energy. As a consequence, they play an important role in low-temperature properties, including thermodynamics and transport. The resulting band is extremely narrow (``flat''), corresponding to a Weyl (or Dirac) dispersion
\begin{equation}
\varepsilon=\hbar v k \label{eq:dispersion}
\end{equation}
with ultralow velocity $v$. $\varepsilon$ and $k$ are the energy and wave vector counted from a Weyl (or Dirac) point. The heat capacity (for a sample of volume $V$) resulting from this dispersion is \cite{Lai18.1}
\begin{equation}
C =\frac{7\pi^2V}{30}k_{\rm B}\left(\frac{k_{\rm B}T}{\hbar v}\right)^3 = \Gamma T^3\; , \label{eq:C}
\end{equation}
which is indeed experimentally observed \cite{Dzs17.1}, as will be shown later.

Furthermore, a magnetic field-tuning experiment \cite{Dzs22.1}, also detailed below, together with theoretical work on the field-tuning effect \cite{Gre20.1x} revealed that, with increasing magnetic field, Weyl nodes and their respective anti-nodes move mostly (for details see \cite{Gre20.1x}) at constant energy in momentum space until they meet and annihilate. The theoretical work considers an Anderson lattice model on a diamond crystal structure with an inversion-symmetry-breaking sublattice potential and is solved in the strong-coupling (Kondo) limit using the auxiliary boson method \cite{Gre20.1x}. Torque magnetization measurements \cite{Dzs22.1} furthermore demonstrated that the Weyl nodes are positioned within a Kondo insulator gap. For Ce$_3$Bi$_4$Pd$_3$, this situation is expected in analogy with the well-known Kondo insulator Ce$_3$Bi$_4$Pt$_3$ \cite{Buc94.1,Hun90.1,Rey94.1,Jai00.1}, which is an isostructural and isoelectronic sibling of Ce$_3$Bi$_4$Pd$_3$ \cite{Dzs17.1,Dzs22.1}. The topological nodal states are situated within the gap because, apparently, they are robust against being gapped out in the Kondo hybridization process \cite{Dzs22.1}. The gapped background, identified also in \cite{Kus19.1}, is a fortuitous situation for experiments because abundant topologically trivial states at the Fermi level might otherwise cover the effect of the topological nodal states.

The key transport signature of a Weyl--Kondo semimetal is the ``spontaneous'' Hall effect \cite{Dzs21.1}. The term spontaneous refers to the situation that a transverse voltage appears in response to a (longitudinal) electrical current but in the absence of both internal and external magnetic fields. An approximate formulation of the Hall response in a time-reversal symmetric but inversion asymmetric setting is
\begin{equation}
j_y = \sigma_{xy} \mathcal{E}_x = \frac{e^3\tau}{\hbar^2} \underbrace{\int \frac{d^3k}{(2\pi)^3} f_0(\bm{k}) \frac{\partial\Omega^{\rm{odd}}_z(\bm{k})}{\partial k_x}}_{D_{xz}}\mathcal{E}_x^2\; , \label{eq:jy}
\end{equation} 
where $\mathcal{E}_x$ is an electric field applied along $x$ (``longitudinal''), $\Omega^{\rm{odd}}$ the Berry curvature, which is odd in momentum space, $f_0(\bm{k})$ the equilibrium Fermi--Dirac distribution function, $D_{xz}$ the Berry curvature dipole, and $j_y$ the resulting transverse (Hall) current density \cite{Sod15.1}. This is the first nonvanishing term in an expansion in the longitudinal electric field. In this limit, the spontaneous Hall conductivity $\sigma_{xy}$ is proportional to $\mathcal{E}_x$; thus this response has also been called ``nonlinear'' Hall effect. In Weyl--Kondo semimetals, however, the Weyl nodes can be situated so close to the Fermi energy that, even for small applied electric fields, higher order terms are needed to capture the experimentally observed behavior \cite{Dzs21.1}. Indeed, in the first candidate material, Ce$_3$Bi$_4$Pd$_3$, which is time reversal invariant as demonstrated by zero-field muon spin rotation ($\mu$SR) experiments \cite{Dzs21.1} but has a noncentrosymmetric crystal structure, not only the square-in-$\mathcal{E}_x$ spontaneous Hall current (or voltage) expected from Eq.\,\ref{eq:jy} but also a contribution that is linear in $\mathcal{E}_x$ was observed and attributed to higher order terms that are neglected in Eq.\,\ref{eq:jy} \cite{Dzs21.1}. In applied magnetic fields (or magnetic induction $B=\mu_0H$), the spontaneous Hall effect finds continuation as an even-in-$B$ Hall response. The magnetic field can be ruled out to be the origin of this effect, as it would necessarily always result in an odd-in-$B$ Hall effect.

\begin{figure}[t!]
    \centering
    \includegraphics[height=0.38\textwidth]{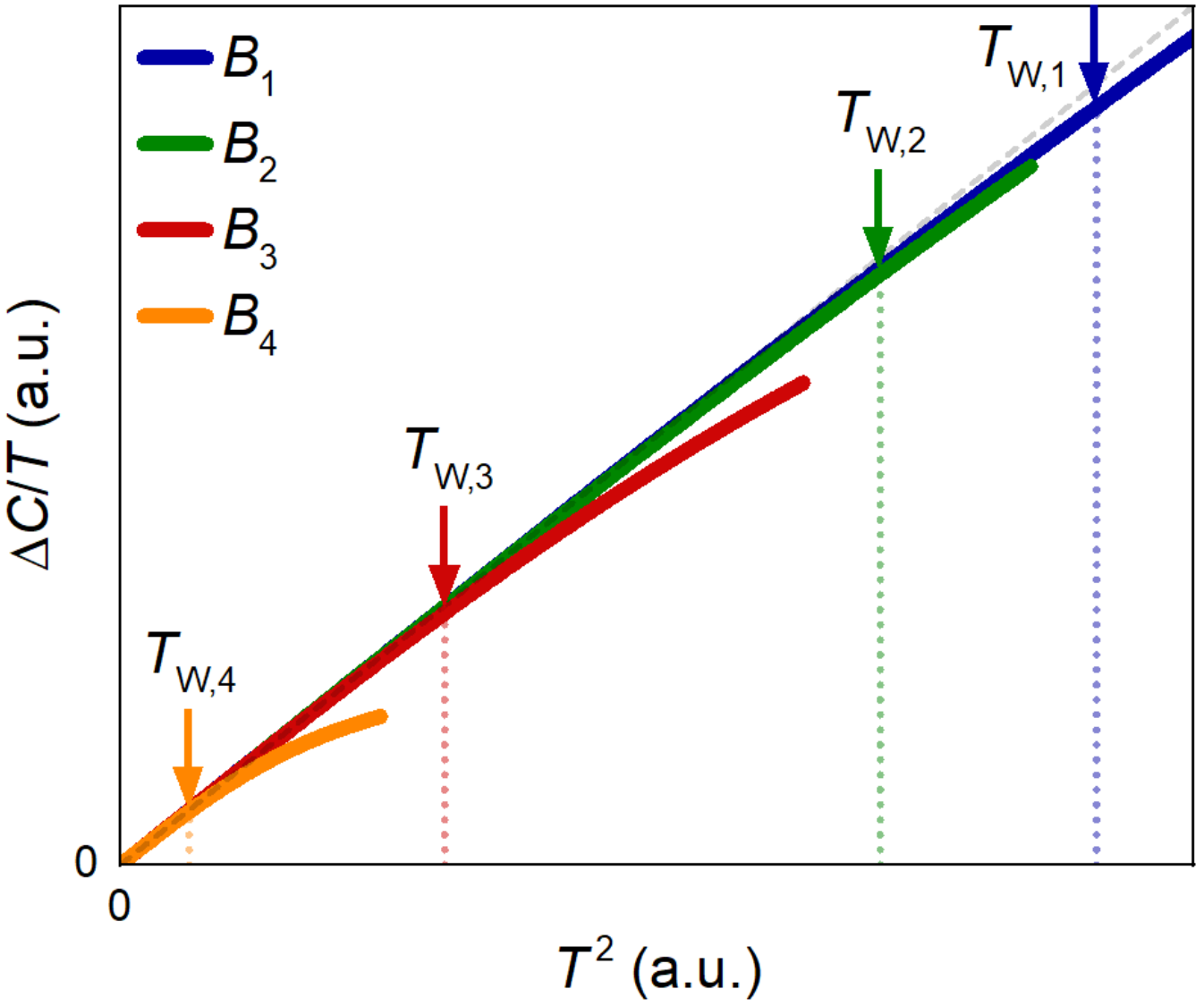}\hspace{0.5cm}\includegraphics[height=0.38\textwidth]{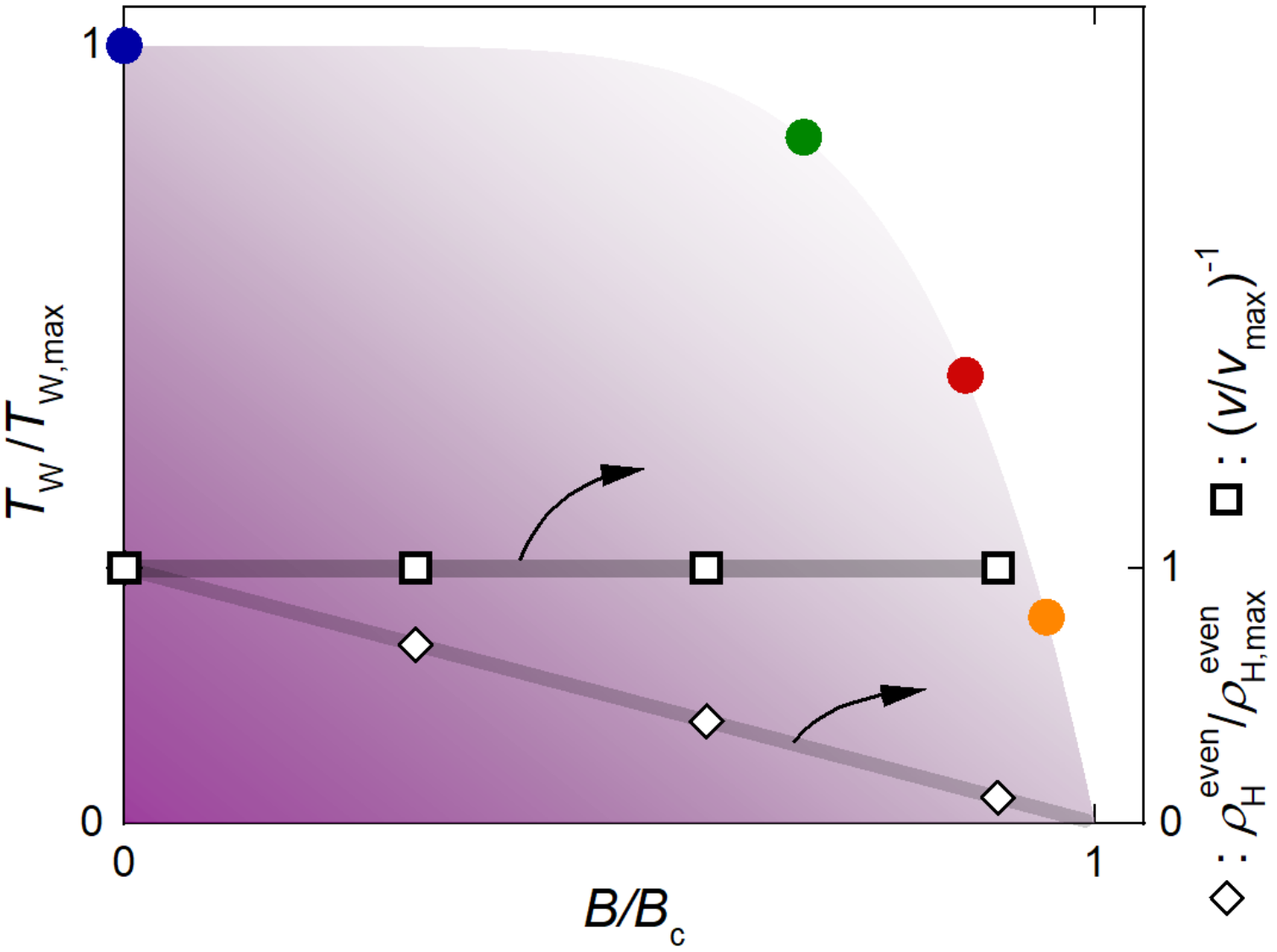}
    \vspace{0.0cm}
    
     \includegraphics[height=0.13\textwidth]{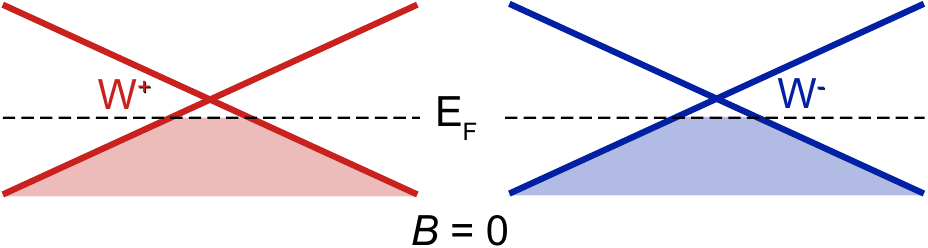}\hspace{1cm}\includegraphics[height=0.13\textwidth]{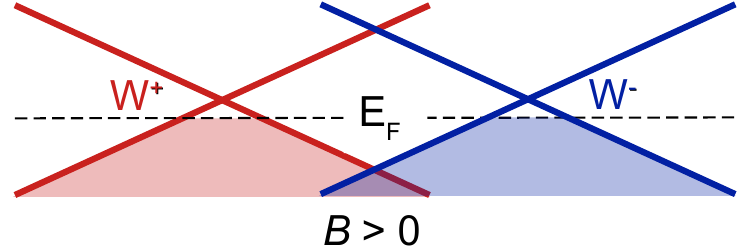}
\vspace{-9.1cm}

\hspace{-8cm}\large\bf\textsf{A} \hspace{6.9cm} \large\bf\textsf{B}
\vspace{5.9cm}

\hspace{-15.6cm}\large\bf\textsf{C}
\vspace{2cm}
 
    \caption{Weyl--Kondo semimetal characteristics. {\bf (A)} Weyl contribution to the heat capacity $\Delta C$, plotted as $\Delta C/T$ vs $T^2$, for different magnetic fields (inductions) $B_i$. The linear behavior, corresponding to a $\Delta C \propto T^3$ dependence, is a thermodynamic signature of bands with linear dispersion (Eq.\,\ref{eq:dispersion}). For Weyl semimetals, its slope is related to the Weyl velocity $v$ via Eq.\,\ref{eq:C}. {\bf (B)} Temperature--magnetic field phase diagram displaying the region (violet shading) in which the Weyl--Kondo semimetal signature in specific heat is observed. $T_{\rm W, max}$ is the temperature up to which the $\Delta C \propto T^3$ dependence holds in zero field. The right axes display the inverse Weyl velocity $1/v$ (squares) and the even-in-field Hall resistivity $\rho_{\rm H}^{\rm even}$ (diamonds), both normalized to their maximum values. {\bf (C)} Sketch of the dispersions near a Weyl (W$^+$) and its anti-Weyl node (W$^-$), in zero magnetic field (left) and in an applied magnetic field (right). The dashed line indicates the Fermi energy $E_{\rm F}$, chosen here to be positioned slightly below the Weyl nodes.}
    \label{fig1}
\end{figure}

In figure \ref{fig1} we sketch these key signatures of a Weyl--Kondo semimetal. According to Eq.\,\ref{eq:C} the Weyl contribution to the heat capacity $\Delta C$ shows linear behavior on a $\Delta C/T$ vs $T^2$ plot (figure \ref{fig1}A), with a slope $\Gamma$ that is inversely proportional to $v^3$. Because the Kondo interaction can lead to bandwidth renormalizations of several orders of magnitude, $v$ will be drastically reduced compared to the Fermi velocity of simple noninteracting Schr\"odinger-like quasiparticles (e.g., $1.4\times 10^6$\,m/s for gold) or, perhaps more significantly, noninteracting Dirac-like quasiparticles (e.g., $1\times 10^6$\,m/s for graphene \cite{Cas09.1}). This reduction of $v$ boosts the heat capacity to the point that it may even overshoot the low-temperature (Debye-like) phonon contribution \cite{Dzs17.1}. The temperature $T_{\rm W}$ up to which this law holds is a measure of the stability of the Weyl--Kondo semimetal phase. It is plotted as full circles in figure \ref{fig1}B. We note that, unlike broken-symmetry phases characterized by an order parameter, this state is not bound by a phase transition but builds up continuously as Kondo coherence sets in \cite{Lai18.1,Dzs21.1}. This is symbolized by the violet shading, which lacks a sharp boundary. With increasing applied magnetic field, $T_{\rm W}$ is successively suppressed. This is because the Weyl and anti-Weyl dispersions start to intersect as the Weyl nodes move towards each other in momentum space \cite{Gre20.1x} (figure \ref{fig1}C). The Weyl--Kondo semimetal phase collapses when the Weyl and anti-Weyl nodes meet and annihilate. The slope of the dispersions is, a priori, not expected to vary with $B$, as visualized by the inverse Weyl velocity plotted as squares in figure \ref{fig1}B on the right $y$ axis. The magnitude of the even-in-$B$ Hall effect, by contrast, depends on the momentum-space distance between a Weyl and the associated anti-Weyl node \cite{Gre20.1x}. As such it is expected to decrease with increasing field (see diamonds plotted on the right $y$ axis in figure \ref{fig1}B).

\section{Weyl--Kondo semimetal candidate materials}\label{cand}
\begin{figure}[t!]
    \centering
    \includegraphics[width=\textwidth]{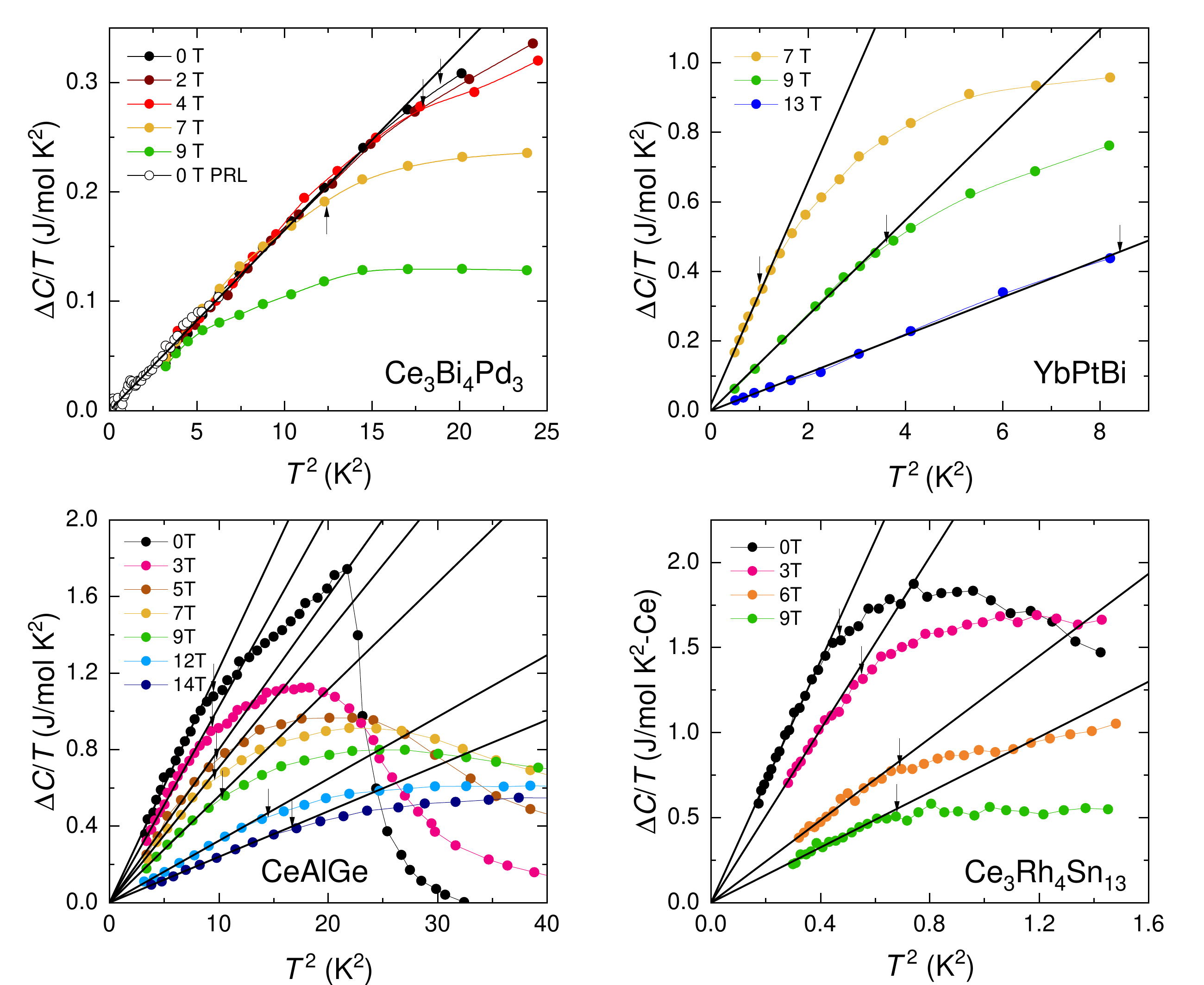}
\vspace{-13.5cm}

\hspace{-7.4cm}\large\bf\textsf{A} \hspace{7.1cm} \large\bf\textsf{B}
\vspace{5.7cm}

\hspace{-7.4cm}\large\bf\textsf{C} \hspace{7.1cm} \large\bf\textsf{D}
\vspace{5.9cm}

    \caption{$\Delta C/T$ vs $T^2$ at fixed magnetic fields for {\bf(A)} Ce$_3$Bi$_4$Pd$_3$, {\bf(B)} YbPtBi, {\bf(C)} CeAlGe, and {\bf(D)} Ce$_3$Rh$_4$Sn$_{13}$. For Ce$_3$Bi$_4$Pd$_3$, $\Delta C/T = C/T - (\gamma + \beta T^2)$, where $C$ is the total measured specific heat, $\gamma$ a zero-temperature offset, which might originate from residual topologically trivial ``background'' bands, and $\beta$ the prefactor of the low-temperature phonon contribution as determined from the non-$f$ reference compounds La$_3$Bi$_4$Pd$_3$ \cite{Dzs22.1} and La$_3$Bi$_4$Pt$_3$ \cite{Dzs17.1}. For the three other compounds, we used $\Delta C/T = C/T - \gamma$, i.e., no phonon contribution was subtracted. However, this is not expected to change the conclusions as in all three cases $\beta$ is much smaller than (less than 2\% of) the measured slopes \cite{Guo18.1,Hod18.1,Iwa23.1}, creating a maximum error of 0.6\% on the extracted putative Weyl velocities. Note that for CeAlGe, $\gamma$ is negative above 5\,T, which is unphysical and indicates that the temperature dependence has to change at lower temperatures. The arrows mark the onset temperature $T_{\rm W}$ of the $\Delta C \propto T^3$ behavior, defined here via a deviation of the data by more than 5\% from the low-temperature $\Delta C/T = \Gamma T^2$ fit. For YbPtBi, the onset temperatures tabulated in \cite{Guo18.1} were taken, where the definition criterion is not further specified. The data for the plots were taken from \cite{Dzs22.1} (Ce$_3$Bi$_4$Pd$_3$, open symbols from Ref.\,\cite{Dzs17.1}), \cite{Guo18.1} (YbPtBi), \cite{Hod18.1} (CeAlGe), and \cite{Iwa23.1} (Ce$_3$Rh$_4$Sn$_{13}$).}
    \label{fig2}
\end{figure}

The above-described experiments on Ce$_3$Bi$_4$Pd$_3$ \cite{Dzs17.1,Dzs21.1} together with the theoretical studies on nonsymmorphic Kondo lattice models \cite{Lai18.1,Gre20.1} have coined the notion of the Weyl--Kondo semimetal. This sets the stage to consider experimental results on other noncentrosymmetric compounds in this context. In what follows we review the pertinent data and compare them with the behavior seen in Ce$_3$Bi$_4$Pd$_3$. In figure \ref{fig2} we replot published specific heat data, in the form of isofield $\Delta C/T$ vs $T^2$ curves, for Ce$_3$Bi$_4$Pd$_3$ \cite{Dzs22.1}, the cubic half-Heusler compound YbPtBi (SG $F\bar{4}3m$, No.\ 216) \cite{Guo18.1}, the tetragonal compound CeAlGe (SG $I4_1md$, No.\ 109) \cite{Hod18.1,Sin20.1,Cor21.1}, and the cubic compound Ce$_3$Rh$_4$Sn$_{13}$ (SG $I2_13$, No.\ 199) \cite{Kuo18.1,Iwa23.1}, in panels A-D, respectively. \red{Details of the data analyses are explained in the caption.}

For all four compounds these plots display ranges of linearity, as expected for a Weyl--Kondo semimetal according to Eq.\,\ref{eq:C}. However, a closer inspection reveals distinct differences from the behavior of Ce$_3$Bi$_4$Pd$_3$. Firstly, the maximum temperature $T_{\rm W}$ up to which the linear behavior holds {\it in}creases with $B$ for YbPtBi, CeAlGe, and Ce$_3$Rh$_4$Sn$_{13}$, whereas it {\it de}creases with $B$ for Ce$_3$Bi$_4$Pd$_3$. This would indicate that, in these other compounds, magnetic field stabilizes the Weyl--Kondo semimetal phase as opposed to the suppression predicted from Zeeman coupling tuning \cite{Gre20.1x}. Secondly, the slopes of the linear dependencies are sizably reduced with $B$, whereas for Ce$_3$Bi$_4$Pd$_3$ all iso-$B$ curves have essentially the same slope. The (putative) Weyl dispersions do thus not remain unchanged (as in the cartoon in figure \ref{fig1}) but become steeper under magnetic field tuning. This sizable correlation tuning effect may hint at the presence of a nearby quantum critical point.

As pointed out in \cite{Dzs17.1}, a $T^3$ contribution to the specific heat may alternatively result from 3D antiferromagnetic (AFM) magnons, as seen for instance in the heavy fermion antiferromagnets CeIn$_3$ \cite{Ber79.1,Cor01.1}, CePd$_2$In (between 3 and 6\,T) \cite{Moc96.1}, or CeGe$_{1.76}$ \cite{Bud14.1} below the respective N\'eel temperatures. A sizable reduction of the slope with increasing magnetic field, as seen in YbPtBi, CeAlGe, and Ce$_3$Rh$_4$Sn$_{13}$, would indeed be expected in this situation \cite{Deo73.1,Map06.1}. In fact, CeAlGe is known to order below 5\,K \cite{Hod18.1}, with a complex structure of predominantly antiferromagnetic nature \cite{Pup20.1}, suggesting that AFM magnons may contribute to the observed $\Delta C \propto T^3$ dependence. For Ce$_3$Rh$_4$Sn$_{13}$, there are conflicting reports on its magnetic order. Whereas in \cite{Odu07.1} two antiferromagnetic phase transitions at 2 and 1.2\,K were reported from specific heat measurements, no clear evidence for magnetic order was found in neutron diffraction experiments \cite{Suy18.1}. This calls for further investigations, for instance by zero-field $\mu$SR experiments, which ruled out even spurious magnetism in Ce$_3$Bi$_4$Pd$_3$ \cite{Dzs21.1}. YbPtBi is known to order antiferromagnetically in zero field, but this order is suppressed to $T=0$ at 0.4\,T and a Fermi liquid state is recovered at fields above 0.8\,T \cite{Mun13.1}. The $\Delta C \propto T^3$ dependence highlighted in \cite{Guo18.1} appears only at a much larger field of 7\,T, deep in the Fermi liquid region. This seems to rule out a connection with the low-field AFM phase. On the other hand, it remains to be understood why a compound with broken inversion symmetry (such as YbPtBi) would not exhibit Weyl--Kondo semimetal features at smaller fields (including $B=0$). It should also be clarified whether the $B$-induced increase of the crystal electric field level splitting evidenced in \cite{Mun13.1} may underlie the strong field dependence of the $\Delta C/T$ data (figure \ref{fig2}B).

\begin{figure}[t!]
    \centering
    
    \includegraphics[width=0.48\textwidth]{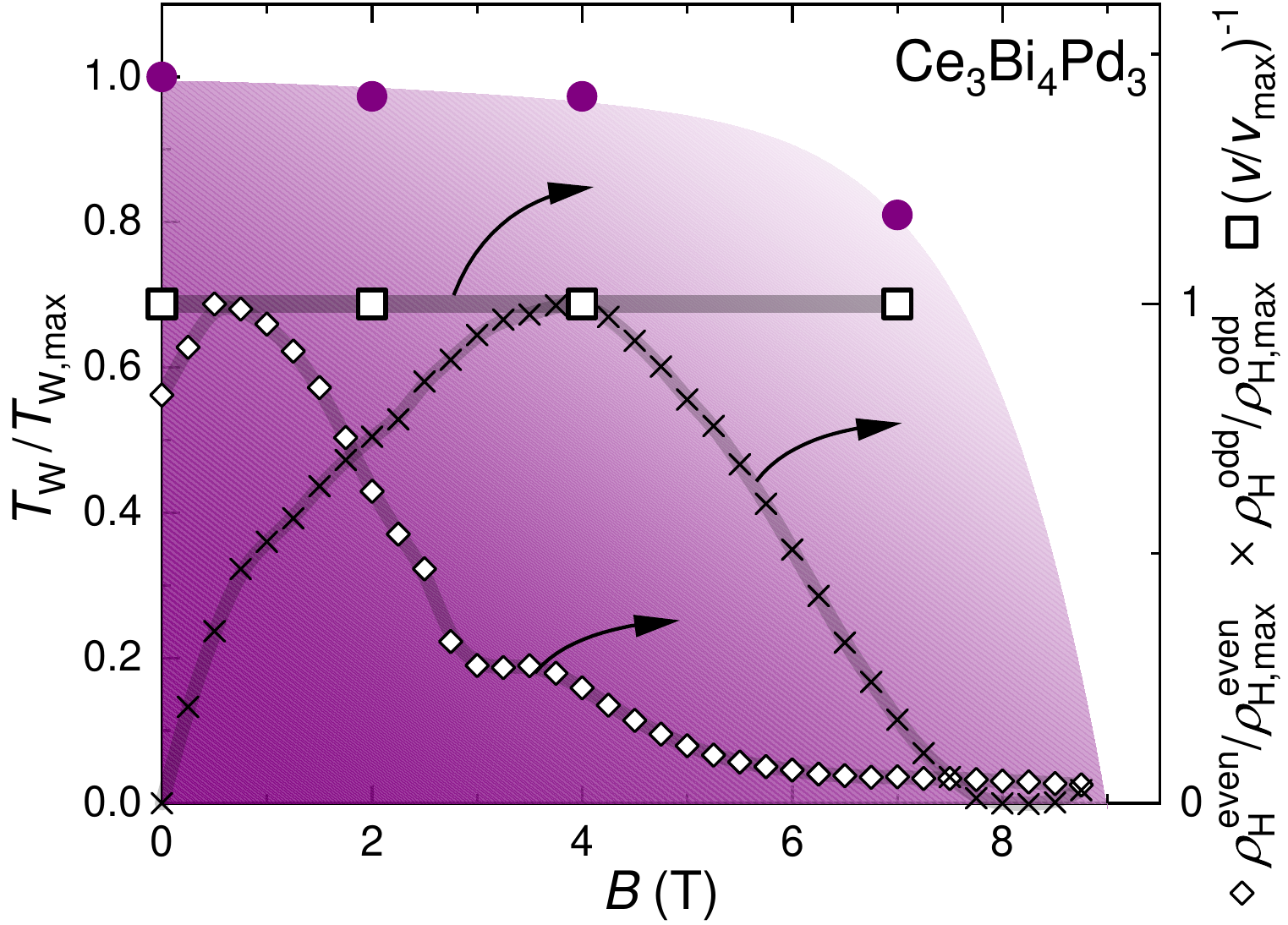} \hspace{0.3cm} \includegraphics[width=0.48\textwidth]{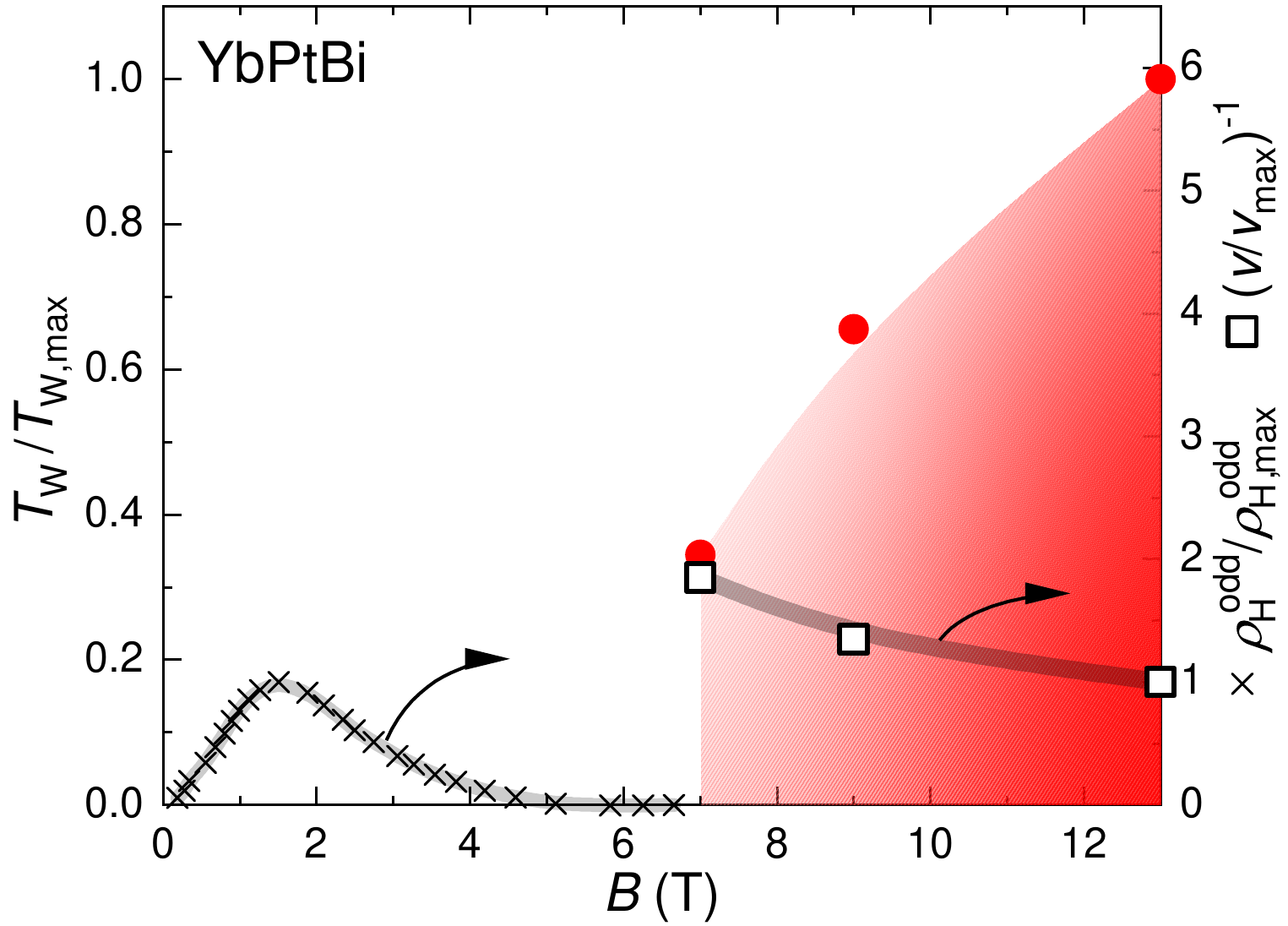}

    \vspace{0.3cm}
    
    \includegraphics[width=0.48\textwidth]{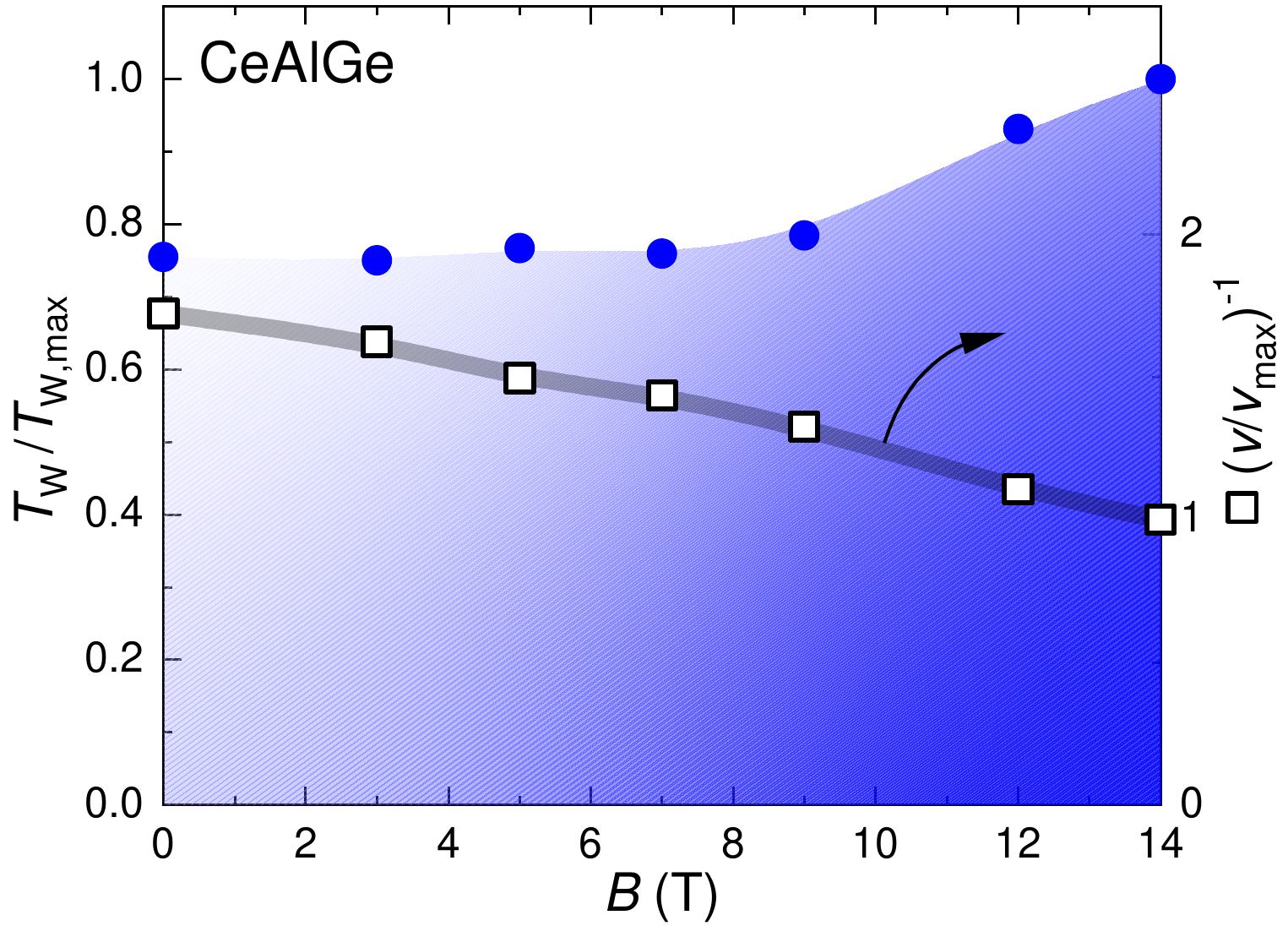} \hspace{0.3cm} \includegraphics[width=0.48\textwidth]{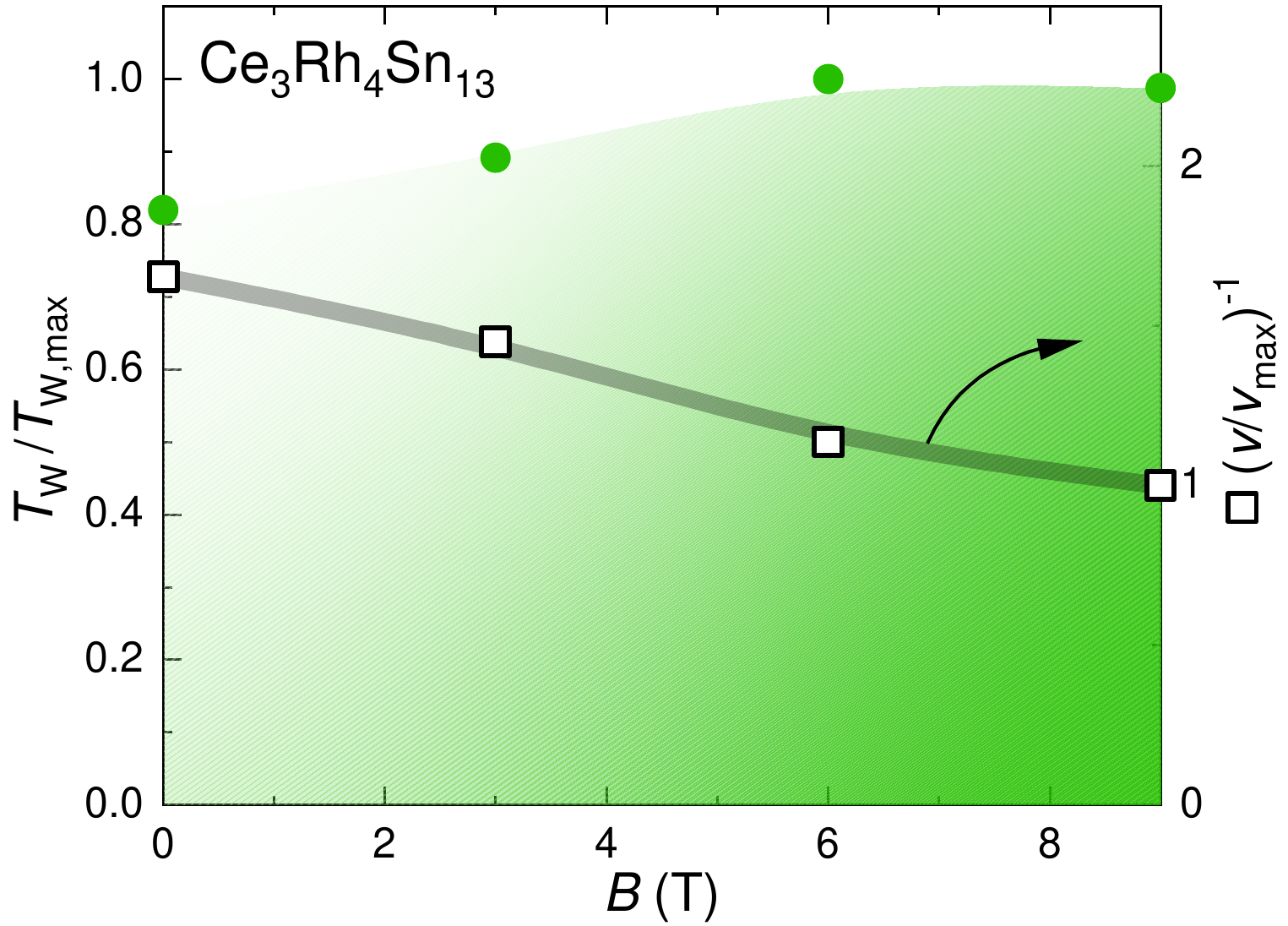}
\vspace{-12.2cm}

\hspace{-7.6cm}\large\bf\textsf{A} \hspace{7.4cm} \large\bf\textsf{B}
\vspace{5.1cm}

\hspace{-7.6cm}\large\bf\textsf{C} \hspace{7.4cm} \large\bf\textsf{D}
\vspace{5.4cm}
   
    \caption{Temperature--magnetic field phase diagrams for {\bf(A)} Ce$_3$Bi$_4$Pd$_3$, {\bf(B)} YbPtBi, {\bf(C)} CeAlGe, and {\bf(D)} Ce$_3$Rh$_4$Sn$_{13}$, for comparison with the expectation for a Weyl--Kondo semimetal sketched in figure \ref{fig1}B. The field-dependent onset temperatures $T_{\rm W}$ (circles, left axes) of the $\Delta C \propto T^3$ behavior, defined as explained in figure \ref{fig2} and normalized by the respective maximum value $T_{\rm W,max}$, delineate the region of (putative) Weyl--Kondo semimetal behavior (shading). The right axis displays the field dependence of the inverse (putative) Weyl velocities $1/v$ extracted from the slopes $\Gamma$ of the linear fits in figure \ref{fig2} (squares) and, where available, the even-in-field Hall resistivity $\rho_{\rm H}^{\rm even}$ (diamonds) and an ``anomalous'' odd-in-field Hall resistivity $\rho_{\rm H}^{\rm odd}$ (crosses), all normalized by the respective maximum values. The (putative) Weyl velocities for Ce$_3$Bi$_4$Pd$_3$ and YbPtBi are taken from \cite{Dzs17.1} and \cite{Guo18.1}, respectively. For CeAlGe and Ce$_3$Rh$_4$Sn$_{13}$, they were calculated in this work using Eq.\,\ref{eq:C}. The Hall data are the lowest-temperature isotherms available, which were taken at 0.4\,K for Ce$_3$Bi$_4$Pd$_3$ \cite{Dzs21.1, Dzs22.1} and 0.3\,K for YbPtBi \cite{Guo18.1}.}
    \label{fig3}
\end{figure}

In figure \ref{fig3} we summarize the characteristics extracted for all four compounds in temperature--magnetic field phase diagrams. As in figure \ref{fig1}B, the full circles represent the onset temperatures $T_{\rm W}$ of the $\Delta C \propto T^3$ behavior and the open squares the (putative) inverse Weyl velocities extracted from the slopes $\Gamma$ of linear fits to $\Delta C/T$ vs $T^2$. For Ce$_3$Bi$_4$Pd$_3$ the Weyl velocity is approximately constant within the magnetic field range where the Weyl--Kondo semimetal exists. For the other three compounds, a pronounced field dependence is observed which, as discussed above, may hint at alternative origins of the $\Delta C \propto T^3$ dependencies.

A spontaneous (nonlinear) Hall effect has so far only been observed for Ce$_3$Bi$_4$Pd$_3$ (diamond at $B=0$ in figure \ref{fig3}A). It is seen as a ``smoking gun'' signature for Weyl nodes in a time reversal symmetric but inversion-symmetry-broken semimetal, as the Berry curvature divergences at the Weyl nodes are its only plausible origin. If they are placed very close to the Fermi energy, as expected in a Weyl--Kondo semimetal \cite{Lai18.1}, the resulting spontaneous Hall effect may be giant. Also the corresponding finite field signature, the even-in-field Hall effect as seen in Ce$_3$Bi$_4$Pd$_3$ \cite{Dzs21.1} (diamonds at $B>0$ in figure \ref{fig3}A), remains to be discovered in the other Weyl--Kondo semimetal candidate materials.

What has been analyzed and proposed as evidence for Weyl physics in YbPtBi is an odd-in-field Hall effect (crosses in figure \ref{fig3}B) \cite{Guo18.1}. It represents a magnetic-field {\em induced} effect, in contrast to the spontaneous Hall effect, which exists in $B=0$, and the even-in-field Hall effect, which exists {\em in spite of} the presence of a finite field (i.e., the field is not its origin). As such, it is more ambiguous evidence for Weyl semimetal physics. In general, the identification of intrinsic Berry curvature contributions in odd-in-field Hall resistivity data is a nontrivial task, which has led to conflicting results in particular in magnetic materials \cite{Tia09.1}. In Ce$_3$Bi$_4$Pd$_3$, such a contribution was identified as the deviation from a linear-in-field Hall resistivity, which is observed only at low temperatures and fields, within the Weyl--Kondo semimetal regime \cite{Dzs21.1} (crosses in figure \ref{fig3}A). Note that in this regime the magnetization is linear in field and can thus not be at the origin of this effect. This contribution is necessarily zero for $B=0$, then increases to its maximum value, and vanishes again as the Weyl--Kondo semimetal is suppressed by magnetic field. For YbPtBi, the odd-in-field Hall signal appears to exist outside the putative Weyl--Kondo semimetal regime identified via the specific heat (red shading in figure \ref{fig3}B), which calls for measurements at lower temperatures to verify whether the putative Weyl--Kondo semimetal regime might persist to lower fields.

\section{Quantifying the correlation strength of Weyl--Kondo semimetals}\label{quant}

The Weyl--Kondo semimetal Ce$_3$Bi$_4$Pd$_3$ was shown to exhibit ``giant'' topological responses \cite{Dzs17.1,Dzs21.1}. This was attributed to the strong bandwidth renormalization via the Kondo effect, which results in a flat Weyl dispersion with very low Weyl velocity. It seems plausible that the Kondo effect leads to similar renormalization effects for both Schr\"odinger and Dirac/Weyl-like quasiparticles. Thus, a comparison between the respective renormalization factors can serve as a consistency check.

\begin{figure}[t!]
    \centering
    \includegraphics[width=0.6\textwidth]{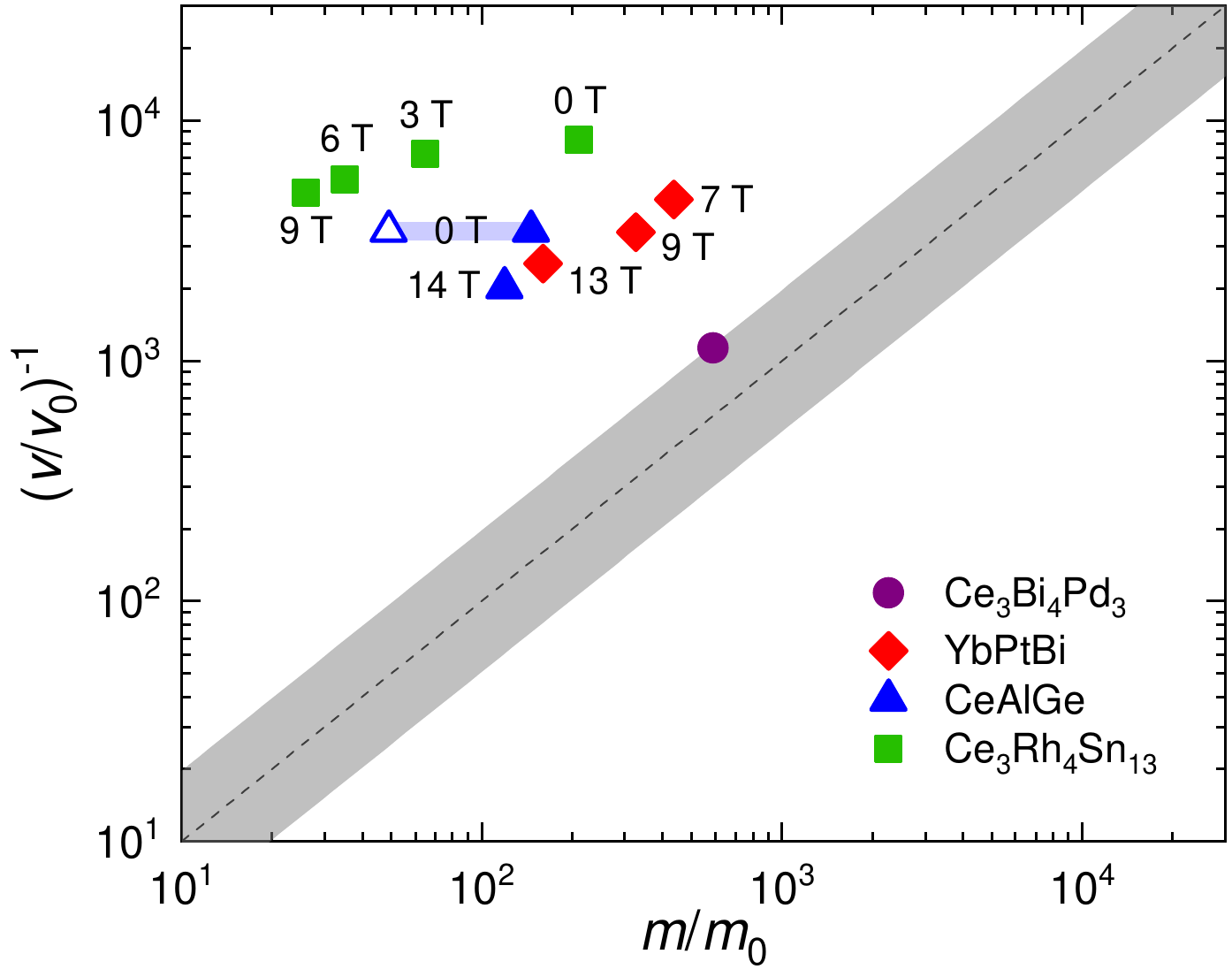}
    \caption{Weyl vs Schr\"odinger renormalization. (Putative) inverse Weyl velocity $v^{-1}$, scaled by the inverse of the Dirac quasiparticle velocity of graphene $v_0 = 1\times 10^6$\,m/s \cite{Cas09.1}, as extracted from the linear-in-$T^3$ electronic (or, more generally, nonphononic) specific heat of Ce$_3$Bi$_4$Pd$_3$, YbPtBi, CeAlGe, and Ce$_3$Rh$_4$Sn$_{13}$ (see figure \ref{fig2}) vs the effective (Schr\"odinger) mass renormalization as calculated via Eq.\,\ref{eq:mren}, using the published Sommerfeld coefficients $\gamma$ and charge carrier concentrations $n$ given in the table \ref{references_fig4}. For the open symbol of CeAlGe, $m/m_0 = 49$ given in \cite{Cor21.1} was used, where it was determined using the plasma frequency instead of $n$.}
    \label{fig4}
\end{figure}

To scrutinize the Weyl--Kondo semimetal interpretation discussed above, we use experimental values of the Sommerfeld coefficient $\gamma$ (removed in the plots in figure \ref{fig2} by plotting $\Delta C/T = C/T - \gamma$) together with Hall effect data for the charge carrier concentration $n$ to estimate the renormalization in the effective (Schr\"odinger) mass via
\begin{equation}
    \frac{m}{m_0} = \frac{3 \hbar^2}{m_0 \cdot k_{\rm B}^2 \cdot (3 \pi^2)^{1/3}} \cdot \frac{\gamma}{n^{1/3}} \; ,
    \label{eq:mren}
\end{equation}
where $m_0$ and $m$ are the free electron mass and the mass renormalized by correlations, respectively, and the other symbols have their usual meaning. As renormalization factor for the Dirac/Weyl quasiparticles, we use $(v/v_0)^{-1}$, i.e.\ the inverse of the (putative) Weyl velocities $v$ from figure \ref{fig3} scaled by $v_0$ (for parameters and references, see table \ref{references_fig4}). The inverse is taken because a larger renormalization of Dirac/Weyl-like bands is reflected by smaller (not larger) velocities. For concreteness, we use $v_0 = 10^6$\,m/s, the Dirac velocity of graphene \cite{Cas09.1}. The expectation for (correlated) Dirac or Weyl semimetals is that $m/m_0$ and $(v/v_0)^{-1}$ have similar values. In the double-logarithmic plot in figure \ref{fig4} this is indicated by a straight line of slope 1. We see that only the data point for Ce$_3$Bi$_4$Pd$_3$ fulfills this expectation.
For the other three materials, the renormalization effect would be much larger for the Dirac/Weyl-like than for the Schr\"odinger-like quasiparticles. This suggests that at least part of the large slopes $\Gamma$ of the $\Delta C \propto T^3$ dependencies of YbPtBi, CeAlGe, and Ce$_3$Rh$_4$Sn$_{13}$ (figure \ref{fig2}) derive from effects other than a Weyl--Kondo semimetal dispersion. In any case, evidence beyond the specific heat signature should be sought to make a Weyl--Kondo semimetal assignment firm.

\begin{table}[h!]
    \centering
    \caption{Parameters used for the data in figure \ref{fig4}, as extracted from the cited publications. The (putative) Weyl velocities $v$ of CeAlGe and Ce$_3$Rh$_4$Sn$_{13}$ were determined within this work from the slopes $\Gamma$ of the linear fits in figure \ref{fig2} using Eq.\,\ref{eq:C}; this is indicated by the $\ast$ after the reference. Because the specific heat of CeAlGe exhibits a phase transition anomaly due to the magnetic ordering, a reliable extraction of $\gamma$ is nontrivial. The values we used at 0\,T and 14\,T correspond to the lowest value of $C/T(B=0)$ above the transition and the lowest measured $C/T(B=14$\,T) value in the entire $T$ range, respectively \cite{Hod18.1}. To obtain the renormalization of the effective mass from Eq.\,\ref{eq:mren} the $\gamma$ values from this table must be converted to SI units (J/(K$^2$m$^3$)) by dividing them by the respective molar volume $V_M$. The carrier density $n$ of Ce$_3$Bi$_4$Pd$_3$ was determined in the region where the Weyl nodes are gapped out (between about 9 and 14\,T) \cite{Dzs22.1}. This is needed for consistency with the Sommerfeld coefficient $\gamma$, which also counts only the Schr\"odinger-like carriers.}
    \vspace{0.3cm}
    \begin{tabular}{|c|c|c|c|}
        \hline
        compound & $v$ (m/s) & $\gamma$ (J/(mol\,K$^2$)) & $n$ (1/cm$^3$) \\
        \hline
        Ce$_3$Bi$_4$Pd$_3$ & 885 \cite{Dzs17.1} & 0.627 \cite{Dzs22.1} & 8.2$\cdot 10^{19}$ \cite{Dzs21.1,Dzs22.1} \\
        
        YbPtBi\,(7\,T) & 213 \cite{Guo18.1} & 0.244 \cite{Guo18.1} & $5.2 \cdot 10^{20}$ \cite{Sch17.1} \\
        YbPtBi\,(9\,T) & 292 \cite{Guo18.1} & 0.182 \cite{Guo18.1} & $5.2 \cdot 10^{20}$ \cite{Sch17.1} \\
        YbPtBi\,(13\,T) & 394 \cite{Guo18.1} & 0.089 \cite{Guo18.1} & $5.2 \cdot 10^{20}$ \cite{Sch17.1} \\
         
        CeAlGe\,(0\,T) & 288 \cite{Hod18.1}$^{\ast}$ & 0.05 \cite{Hod18.1} &  $1.4 \cdot 10^{20}$ \cite{Hod18.1} \\
        CeAlGe\,(14\,T) & 496 \cite{Hod18.1}$^{\ast}$ & 0.041 \cite{Hod18.1} & $1.4 \cdot 10^{20}$ \cite{Hod18.1} \\
        
        Ce$_3$Rh$_4$Sn$_{13}$\,(0\,T) & 121 \cite{Iwa23.1}$^{\ast}$ & 3.44 \cite{Iwa23.1} & $5 \cdot 10^{22}$ \cite{Koe07.2} \\
        Ce$_3$Rh$_4$Sn$_{13}$\,(3\,T) & 138 \cite{Iwa23.1}$^{\ast}$ & 1.06 \cite{Iwa23.1} & $5 \cdot 10^{22}$ \cite{Koe07.2} \\
        Ce$_3$Rh$_4$Sn$_{13}$\,(6\,T) & 176 \cite{Iwa23.1}$^{\ast}$ & 0.57 \cite{Iwa23.1} & $5 \cdot 10^{22}$ \cite{Koe07.2} \\
        Ce$_3$Rh$_4$Sn$_{13}$\,(9\,T) & 200 \cite{Iwa23.1}$^{\ast}$ & 0.43 \cite{Iwa23.1} & $5 \cdot 10^{22}$ \cite{Koe07.2} \\
         \hline  
    \end{tabular}
    \label{references_fig4}
\end{table}

\section{Topological response vs correlation strength}\label{topo}

As discussed above, the giant spontaneous Hall effect of Ce$_3$Bi$_4$Pd$_3$ may represent such firm evidence. To the best of our knowledge, it has so far not been reported in any other strongly correlated nonmagnetic (time-reversal-symmetry-preserving) Weyl semimetal candidate material, including the three above-discussed heavy fermion compounds. To nevertheless examine whether its magnitude depends on the correlation strength, we resort to a comparison with noninteracting/weakly interacting reference materials. In studies of these compounds, the term nonlinear Hall effect (NLHE) is used, and reference is made to the Berry curvature dipole $D_{xz}$ (see Eq.\,\ref{eq:jy}). As explained in section \ref{char}, this is only part of the \red{Berry-curvature-related} response observed in Ce$_3$Bi$_4$Pd$_3$. Further terms arise when expanding the out-of-equilibrium distribution function around a finite-current setpoint (instead of around $j_x = 0$ as done to obtain Eq.\,\ref{eq:jy}), which is deemed necessary \red{in Weyl--Kondo semimetals} \cite{Dzs21.1}. To discriminate this fully nonequilibrium response from the Berry curvature \red{\it{dipole}} effect (the lowest-order term), the expression ``spontaneous Hall effect'' was used instead of ``NLHE'' \cite{Dzs21.1}. For the purpose of comparison, we adopt the NLHE terminology in what follows.

NLHE studies have been carried out in various (non- or weakly interacting) materials \cite{Kan19.1,Kum21.1,Ma19.1,He21.1,He22.1,Ho21.1,Tiw21.1,Dua22.1,Zha22.2,Min23.1}, but the identification of intrinsic Berry curvature contributions has been challenging. It involves the separation from extrinsic contributions due to effects such as side jump and skew scattering \cite{Du19.1}. For Ce$_3$Bi$_4$Pd$_3$, the Hall angle is constant in the Weyl--Kondo semimetal regime, i.e.,
\begin{equation}
    \tan\Theta = \frac{\sigma_{xy}}{\sigma_{xx}} = \mbox{const} \; , \label{eq:NLHE0}
\end{equation}
as seen from the approximate linear $\sigma_{xy}$ vs $\sigma_{xx}$ dependence below about 3\,K (Fig.\,2B in \cite{Dzs21.1}). Interestingly, this holds for both the dc (and, by extension, the $2\omega$) response and the (fully out-of-equilibrium) $1\omega$ response. In the context of the NLHE, only the $2\omega$ response is considered and it is investigated how $\tan\Theta$, typically scaled by the applied longitudinal electric field $\mathcal{E}_x$, depends on the longitudinal conductivity, i.e.
\begin{equation}
    \frac{\tan\Theta}{\mathcal{E}_x^{\omega}} = \frac{\sigma_{xy}}{\sigma_{xx} \mathcal{E}_x^{\omega}} = \frac{\mathcal{E}_{xy}^{2\omega}}{(\mathcal{E}_x^{\omega})^2} = f(\sigma_{xx}) \; . \label{eq:NLHE00}
\end{equation}
The influence of disorder scattering was studied in a 2D tilted massive (gapped) Dirac model as a minimal symmetry-allowed model for a NLHE \cite{Du19.1}. $f(\sigma_{xx}) = \mbox{const}$, as observed for Ce$_3$Bi$_4$Pd$_3$, was found only for the intrinsic contribution (due to the Berry curvature dipole), and side-jump and skew-scattering terms from dynamic (e.g. phonon-induced) disorder. These latter should depend on temperature and disappear in the zero-temperature limit. The fact that, for Ce$_3$Bi$_4$Pd$_3$, $f(\sigma_{xx}) = \mbox{const}$ holds over the entire temperature range of Weyl--Kondo semimetal behavior is strong evidence for dynamic disorder effects playing a minor role and thus for the intrinsic nature of the spontaneous (or nonlinear) Hall effect. We note that also the linear-response anomalous Hall effect from skew scattering and side-jump scattering was shown to be negligibly small in Ce$_3$Bi$_4$Pd$_3$ (see SI, part B of \cite{Dzs21.1}). As extrinsic scattering effects in the linear-response and nonlinear regimes are related \cite{Kan19.1,Du19.1,Xia19.1}, this is a further confirmation for their absence in the NLHE in Ce$_3$Bi$_4$Pd$_3$. In general, the situation is considerably more complex. Here we focus on investigations of \red{(Pb$_{1-x}$Sn$_x$)$_{1-y}$In$_y$Te \cite{Zha22.2},} MoTe$_2$ \cite{Tiw21.1}, WTe$_2$ \cite{Kan19.1}, and TaIrTe$_2$ \cite{Kum21.1}, \red{where---via stoichiometry optimization in (Pb$_{1-x}$Sn$_x$)$_{1-y}$In$_y$Te to reach a ferroelectric state with extremely low carrier concentration and via exfoliation in the other three noninteracting/weakly interacting reference compounds---a Berry curvature dipole contribution to the NLHE became sufficiently large to be identified with some confidence.}

\red{(Pb$_{1-x}$Sn$_x$)$_{1-y}$In$_y$Te is an In-doped alloy of two rock salt-type compounds: the normal insulator (NI) PbTe and the topological crystalline insulator SnTe. For certain compositions ($x$ and $y$ values), ferroelectric order appears, which breaks the inversion symmetry of the undeformed system (SG $Fm\bar{3}m$, No.\,225), thereby enabling the formation of a Weyl semimetal \cite{Zha21.1}. An ``optimally doped'' sample shows an electrical conductivity that decreases with decreasing temperature \cite{Zha22.2}. Using this and the temperature dependent $\mathcal{E}_{xy}^{2\omega,\rm{intr}}/(\mathcal{E}_x^{\omega})^2$, $f(\sigma_{xx})$ can be obtained. An extrapolation of the low-$\sigma_{xx}$ (low-$T$) values to $\sigma_{xx} = 0$ leads to an intrinsic NLHE of $4.35\times 10^{-4}$\,m/V.} Bulk MoTe$_2$ crystallizes in the noncentrosymmetric $T_{\rm d}$-MoTe$_2$ structure of SG $Pmn2_1$ (No.\ 31) \cite{Sun15.1,Qi16.1}, but the exfoliated films of interest here have a lower $Pm$ symmetry \cite{Son16.1}. $f(\sigma_{xx})$ shows a pronounced dependence on $\sigma_{xx}$, with a functional form that changes with temperature. There is also a pronounced thickness dependence. Thinner films have larger residual resistivity (due to surface scattering), which tips the balance between different (extrinsic) scattering processes. The best estimate of the intrinsic Berry curvature contribution comes from the thinnest samples because they have the smallest conductivity and thus the lowest skew-scattering contribution (which is the dominant extrinsic scattering effect at high temperatures). The extrapolation of $f(\sigma_{xx}) \propto \sigma_{xx}^2$ to $\sigma_{xx} = 0$ (at $T\rightarrow \infty$) gives $1.2\times 10^{-6}$\,m/V. This is one order of magnitude larger than the upper bound estimated from DFT calculations of the Berry curvature dipole, so presumably it is still dominated by extrinsic scattering \cite{Tiw21.1}. The situation is similar in $T_{\rm d}$-WTe$_2$ \cite{Kan19.1}. Its SG $Pmn2_1$ \cite{Bro66.1} is again reduced to $Pm$ in exfoliated multilayer films \cite{Son16.1}. For three films of 5-6 layer thickness, $f(\sigma_{xx})$ was found to be proportional to $\sigma_{xx}^2$ in temperature ranges between 2 and 100\,K. Again, the Berry curvature dipole contribution is estimated by extrapolating this dependence to $\sigma_{xx}=0$. That the values obtained for the three films vary by almost an order of magnitude ($0.15 - 1 \times 10^{-9}$\,m/V) is attributed to the different carrier concentrations and mobilities, though no systematic dependence is seen. Finally, also exfoliated samples of $T_{\rm d}$-TaIrTe$_2$ (again SG $Pmn2_1$ for bulk) reveal such behavior \cite{Kum21.1}. Using the same procedure for the thinnest and thus most resistive film yields $1.8 \times 10^{-8}$\,m/V as an estimate for the intrinsic Berry curvature dipole contribution to the NLHE.

\begin{figure}[t!]
    \centering

    \includegraphics[width=0.48\textwidth]{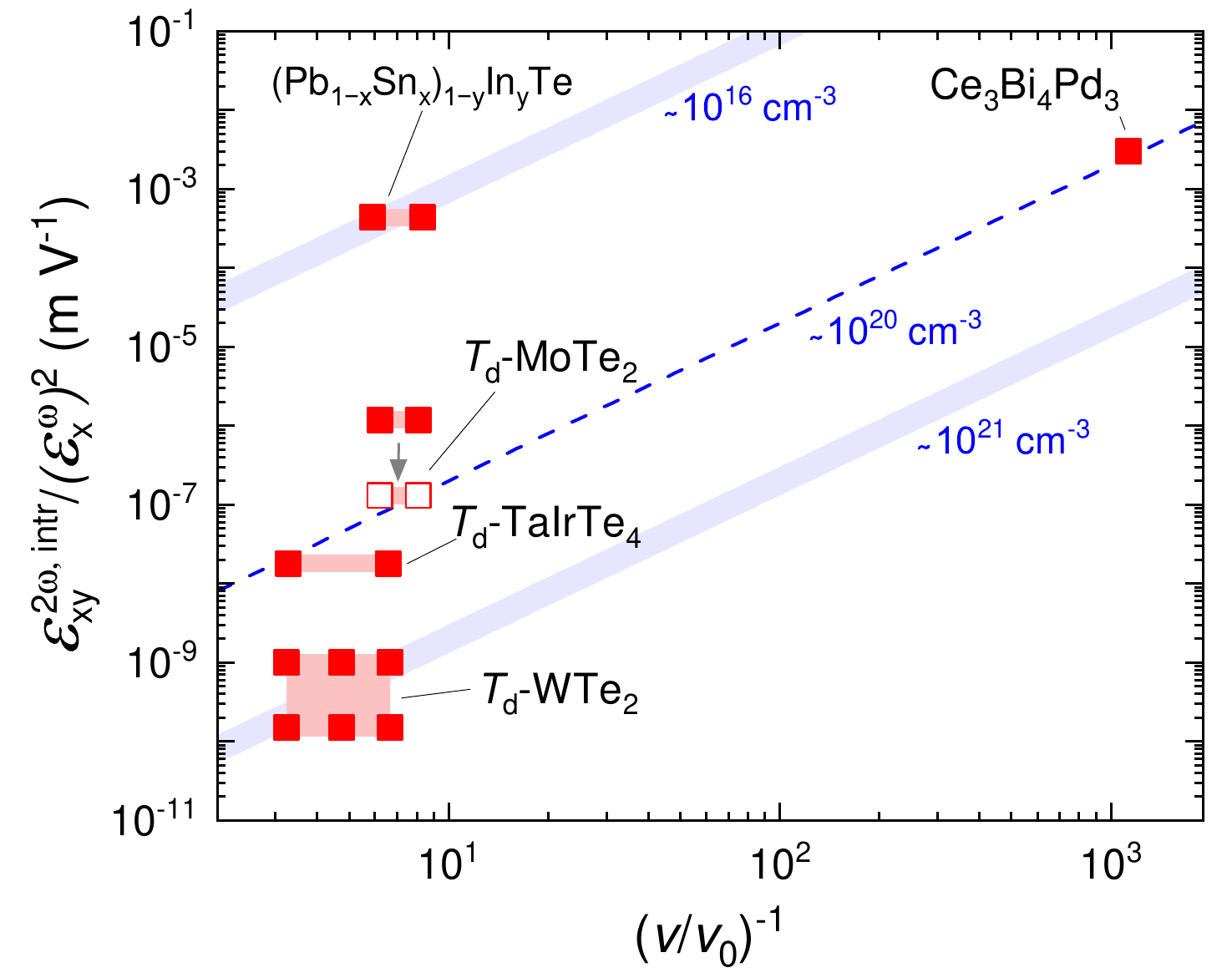}
    \includegraphics[width=0.48\textwidth]{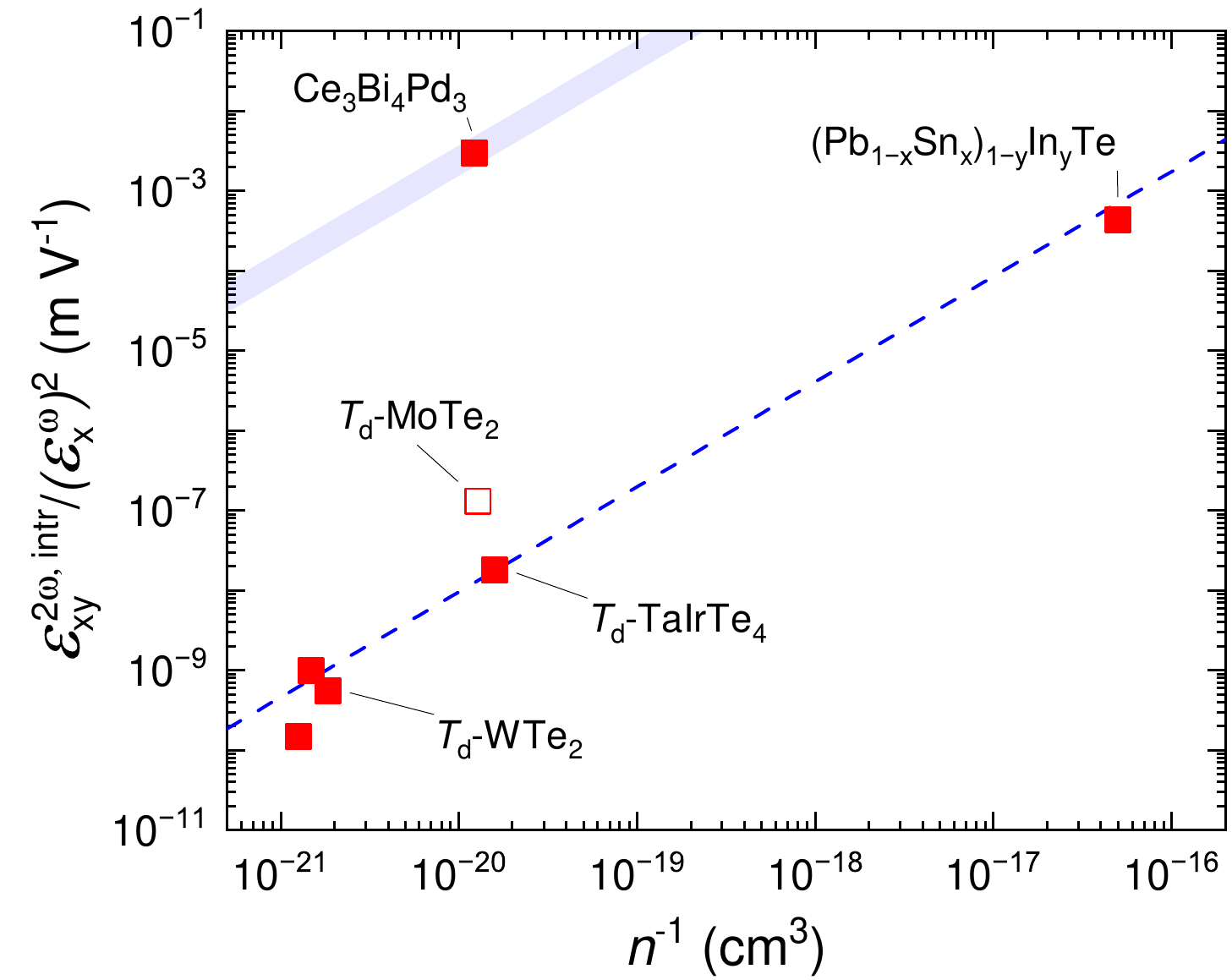}\\

\vspace{-6.2cm}

\hspace{-7.0cm}\large\bf\textsf{A} \hspace{7.1cm} \large\bf\textsf{B}
\vspace{5.2cm}
    \caption{\red{Intrinsic Berry-curvature-dipole-induced nonlinear Hall effect (NLHE), quantified by $\mathcal{E}_{xy}^{2\omega,\rm{intr}}/(\mathcal{E}_x^{\omega})^2$, of (candidate) Weyl semimetals as function of \textbf{(A)} the inverse scaled Weyl velocity $(v/v_0)^{-1}$ as a measure of correlation strength. The full symbols show the experimentally extracted values, the open symbols DFT results. $T_{\rm d}$-TaIrTe$_4$, $T_{\rm d}$-MoTe$_2$, and Ce$_3$Bi$_4$Pd$_3$, which have similar charge carrier concentrations (of order $10^{20}$\,cm$^{-3}$), lie on a universal curve $\sim (v/v_0)^{-2}$, as seen from the dashed guide-to-the-eyes curve. For the two shaded lines at higher and lower $n$, we used the same slope; \textbf{(B)} the inverse charge carrier concentration $n^{-1}$ (of hole-like charge carriers for consistency with \cite{Zha22.2}). The charge carrier concentrations of the quasi-2D material $T_{\rm d}$-WTe$_2$ were calculated using $n=n_{\rm{2D}}/d$, where $d$ is the interlayer distance of (2.7-2.8)\,\AA\ \cite{Son16.1}. The noninteracting/weakly interacting materials lie on a universal curve $\sim n^{-1.3}$, which was determined by fitting (dashed line). A curve with the same slope is plotted through the data point of Ce$_3$Bi$_4$Pd$_3$ (shaded line).} }
    \label{fig5}
\end{figure}

\red{In figure \ref{fig5} we compare the magnitudes of these intrinsic Berry curvature dipole contributions to the NLHE by plotting $\mathcal{E}_{xy}^{2\omega,\rm{intr}}/(\mathcal{E}_x^{\omega})^2$ as a function of the respective reciprocal Weyl velocities $v^{-1}$ (scaled by $v_0^{-1}$, panel A) and charge carrier concentrations $n^{-1}$ (panel B). All values are also given in table \ref{ref_NLAHE}, and the caption contains details on how they were obtained. The data points in figure \ref{fig5}A fall into three groups of similar $n$. In particular, Ce$_3$Bi$_4$Pd$_3$ has roughly the same $n$ as $T_{\rm d}$-TaIrTe$_4$ and $T_{\rm d}$-MoTe$_2$ ($\sim$\,$10^{20}$\,cm$^{-3}$), which is highlighted by the dashed guide-to-the-eyes line which represents a $v^{-2}$ dependence (for the other data points, shaded lines with the same slope are plotted). At constant $n$, $\mathcal{E}_{xy}^{2\omega,\rm{intr}}/(\mathcal{E}_x^{\omega})^2$ thus appears to be boosted by strong correlations, which flatten the Weyl bands (smaller slope $v$ of the Weyl dispersion, Eq.\,\ref{eq:dispersion}) and enhance the electronic density of states at the Fermi level [which scales as $D(E_{\rm F}) \sim v^{-3}$ in a 3D material with Weyl dispersion]. A second trend that becomes clear from this plot is that, at constant $v$, $\mathcal{E}_{xy}^{2\omega,\rm{intr}}/(\mathcal{E}_x^{\omega})^2$ is enhanced by reducing $n$. This dependence is explicitly revealed in figure \ref{fig5}B. All data of the noninteracting/weakly interacting Weyl semimetals fall on a universal curve, $\mathcal{E}_{xy}^{2\omega,\rm{intr}}/(\mathcal{E}_x^{\omega})^2 \sim n^{-1.3}$ (dashed line), evidencing a strong dependence on the proximity of the Weyl nodes to the Fermi energy ($E_{\rm F} \sim n^{1/3}$ in a 3D material with Weyl dispersion). Ce$_3$Bi$_4$Pd$_3$ lies orders of magnitude above this line. Again, we also include a line of the same slope for Ce$_3$Bi$_4$Pd$_3$ (shaded line), which makes a strategy for further enhancing $\mathcal{E}_{xy}^{2\omega,\rm{intr}}/(\mathcal{E}_x^{\omega})^2$ explicit: to reduce the charge carrier concentration in a strongly correlated Weyl semimetal such as Ce$_3$Bi$_4$Pd$_3$. Whether, at constant $n$, the correlation-induced $v$ reduction is the only cause of the drastic enhancement of $\mathcal{E}_{xy}^{2\omega,\rm{intr}}/(\mathcal{E}_x^{\omega})^2$ or whether also other ingredients---such as the multiplicity of Weyl nodes near the Fermi energy, the $k$ space separation of node and anti-node, or the tilting of the nodes \cite{Gre20.1}---contribute, should be clarified by future work.}

\begin{table}[t]
\centering
\caption{Estimates of the intrinsic NLHE contribution due to the Berry curvature dipole, $\mathcal{E}_{xy}^{2\omega,\rm{intr}}/(\mathcal{E}_x^{\omega})^2$ (obtained as explained in the text), \red{the 3D/quasi-2D charge carrier concentration of hole-like charge carriers, $n_{\rm (h)}$,} and the inverse ratio of the Weyl velocity $v$ and the velocity of graphene (taken as $v_0 = 1 \times 10 ^6$\,m/s), for the selected Weyl semimetal (candidate) materials. These data are used in figure \ref{fig5}. The values of $v$ were obtained as follows. Ce$_3$Bi$_4$Pd$_3$: from the slope $\Gamma$ of $\Delta C/T$ vs $T^2$, with the phonon contribution subtracted \cite{Dzs17.1}. \red{(Pb$_{1-x}$Sn$_x$)$_{1-y}$In$_y$Te: from a linear fit of the optical conductivity vs photon energy, yielding $v = 1.7\times 10^5$\,m/s \cite{Zha21.1}, and from a linear-in-$T$ fit of the electrical conductivity, leading to $v = 1.2\times 10^5$\,m/s \cite{Zha22.2}.} MoTe$_2$: from $1/v = m/[\hbar(3\pi^2 n)^{1/3}]$, using the charge carrier concentrations $n = 0.70 \times 10^{20}$ and $0.93 \times 10^{20}$\,1/cm$^3$ of two orbits in Shubnikov--de Haas (SdH) oscillations, and effective masses of $m = (1.0-1.2) m_0$ extracted in a Liftshitz--Kosevich analysis \cite{Zho18.1}. WTe$_2$: from slopes of linearly dispersing (surface) bands in ARPES \cite{Wu16.3}, giving $(1.5-2.1)\times 10^5$\,m/s and from a Weyl orbit in SdH oscillations \cite{Li17.1} on 14-layer thick exfoliated WTe$_2$, giving $v = 3.09 \times 10^5$\,m/s. TaIrTe$_4$: from ARPES revealing linearly dispersing surface states with a slope of 2\,eV\AA\ (or 1\,eV\AA\ as given in the text) \cite{Hau17.2}, yielding a Dirac/Weyl velocity of $v = 3.04\times 10^5$\,m/s (or $1.52\times 10^5$\,m/s).}
\vspace{0.2cm}

\begin{tabular}{|l|l|l|l|}
\hline
Compound & $\mathcal{E}_{xy}^{2\omega,\rm{intr}}/(\mathcal{E}_x^{\omega})^2$ (m/V) & $n_{(h)}$ & $(v/v_0)^{-1}$  \\ \hline
Ce$_3$Bi$_4$Pd$_3$      & 3E-3 \cite{Dzs21.1}	& 	8.2E19	(cm$^{-3}$) \cite{Dzs21.1, Dzs22.1}       & 1130 \cite{Dzs17.1}		\\
\red{(Pb$_{1-x}$Sn$_x$)$_{1-y}$In$_y$Te} & \red{4.35E-4 \cite{Zha22.2}} & \red{2E16 (cm$^{-3}$) \cite{Zha22.2}} & \red{5.9 \cite{Zha21.1}, 8.3 \cite{Zha22.2}} \\
$T_{\rm d}$-MoTe$_2$    & 1.2E-6 \cite{Tiw21.1}  & 7.7E19 (cm$^{-3}$)	\cite{Tiw21.1}  & 6.2-8.1 \cite{Zho18.1}  	 \\
$T_{\rm d}$-WTe$_2$     & (0.15-1)E-9 \cite{Kan19.1}   & (1.49-2.18)E13 (cm$^{-2}$)	\cite{Kan19.1} & 3.2 \cite{Li17.1}, 4.8-6.7 \cite{Wu16.3}  \\
$T_{\rm d}$-TaIrTe$_4$  & 1.8E-8 \cite{Kum21.1}   & 6.3E19 (cm$^{-3}$)	\cite{Kum21.1}    & 3.3-6.6 \cite{Hau17.2}  		  \\  \hline
\end{tabular}
\label{ref_NLAHE}
\end{table}

\section{Discussion and outlook}\label{outlook}
We have investigated the role of strong correlations in topological semimetals. As a starting point, we used the recently discovered time-reversal-invariant but inversion-symmetry-broken Weyl--Kondo semimetal Ce$_3$Bi$_4$Pd$_3$ \cite{Dzs17.1,Lai18.1,Gre20.1,Dzs21.1}. We reviewed its topological signatures in both thermodynamic and transport measurements, namely (i) a ``giant'' value of the electronic specific heat coefficient $\Gamma=\Delta C/T^3$ of Dirac-like quasiparticles, which is associated with ultraslow quasiparticle velocities $v \propto \Gamma^{-1/3}$ and thus ultraflat linearly-dispersing bands \cite{Dzs17.1}; (ii) an equally giant value of the intrinsic nonlinear (spontaneous, i.e.\ $B=0$) Hall effect arising from the Berry curvature monopoles at the Weyl nodes \cite{Dzs21.1}; (iii) a continuation of this zero-field Hall effect as an even-in-$B$ component, confirming that magnetic field is not the cause of the effect \cite{Dzs21.1}; (iv) a clearly identified odd-in-$B$ anomalous Hall effect due to the Berry curvature induced by a magnetic field \cite{Dzs21.1}. We have explained the understanding of these effects in terms of a Weyl--Kondo semimetal model \cite{Lai18.1,Gre20.1} where, at the appropriate filling, Weyl nodes appear in the immediate vicinity of the Fermi level and are associated with Weyl bands with ultraflat dispersion \cite{Dzs17.1,Lai18.1,Gre20.1,Dzs21.1}.

We have produced a temperature--magnetic field phase diagram that delineates the region of Weyl--Kondo semimetal signatures, using the (high-temperature) onset temperature $T_{\rm W}$ of the $\Delta C \propto T^3$ dependence as the ``phase'' boundary (note that a Weyl semimetal is not a phase in the thermodynamic sense). With increasing field, this boundary is suppressed to zero at a critical field $B_{\rm c}$, which is understood in terms of a Zeeman-coupling induced motion of Weyl nodes in momentum space until a Weyl and its anti-Weyl node meet and annihilate \cite{Dzs22.1,Gre20.1x}. We have also included the magnitudes of the topological signatures (i)-(iv), scaled to their maximum values, in this phase diagram. Whereas $\Gamma$ remains essentially constant within the boundary, the Hall signatures get successively suppressed towards $B_{\rm c}$. This behavior indicates that, with increasing field, the Weyl nodes move at constant energy in momentum space, without an appreciable change of the slope of the Weyl bands, until they meet and annihilate at $B_{\rm c}$ \cite{Dzs22.1}, in good agreement with theoretical expectations \cite{Gre20.1x}.

The key aspects that make the Weyl--Kondo semimetal Ce$_3$Bi$_4$Pd$_3$ a prime example for correlation-driven metallic topology are summarized as follows:
\begin{itemize}
    \item Its Weyl--Kondo semimetal phase is well delineated: It emerges only at low temperatures as the material becomes fully Kondo coherent, and is suppressed at a readily accessible magnetic field as the Weyl nodes annihilate.
    \item Its Weyl--Kondo bands reside within a Kondo insulating gap: This eliminates contributions from topologically trivial ``background'' bands to a large extent, aiding the identification of topological signatures; in addition, it pins the Fermi level to the immediate vicinity of the Weyl nodes.
    \item Its Weyl--Kondo semimetal signatures are ``giant'': The orders of magnitude mass renormalization of Schr\"odinger-like quasiparticles known from heavy fermion compounds is inherited by the Weyl quasiparticles in terms of a corresponding band flattening, Weyl velocity suppression, and Weyl density of states enhancement.
\end{itemize}

We have searched the literature for other candidate Weyl--Kondo semimetals and considered the noncentrosymmetric compounds YbPtBi, CeAlGe, and Ce$_3$Rh$_4$Sn$_{13}$ as promising candidates because they all exhibit temperature and field ranges with $\Delta C/T \propto T^2$ behavior with large slopes \cite{Guo18.1,Hod18.1,Iwa23.1}. The phase diagrams that we constructed, however, show several differences from the one of Ce$_3$Bi$_4$Pd$_3$, namely: (i) the (putative) phase boundaries are stabilized as opposed to suppressed with increasing magnetic field; (ii) the (putative) Weyl velocities are significantly increased with field as opposed to essentially unchanged in Ce$_3$Bi$_4$Pd$_3$; (iii) no spontaneous or even-in-field Hall effect is detected; (iv) the odd-in-field Hall effect detected in one of the materials \cite{Guo18.1} seems to appear outside the (putative) phase boundary. As a further consistency check, we estimated effective (Schr\"odinger) masses and Dirac (or Weyl) velocities of the candidate materials. Whereas the expected renormalization ratio of order unity was found for Ce$_3$Bi$_4$Pd$_3$, much stronger Weyl than Schr\"odinger renormalizations would have to be at play in the three other compounds. This calls for further studies, to pin down whether and to which extent other effects (e.g., antiferromagnetic magnons of CEF splitting in large fields) intervene.

Finally, as Ce$_3$Bi$_4$Pd$_3$ is so far the only Weyl--Kondo semimetal in which a spontaneous or even-in-field Hall response has been identified, we resorted to a comparison with noninteracting systems. To the best of our knowledge, the only candidate time-reversal symmetric noninteracting Weyl semimetals that have shown evidence for an intrinsic (Berry-curvature-related) NLHE are exfoliated thin films of $T_{\rm d}$-MoTe$_2$ \cite{Tiw21.1}, $T_{\rm d}$-WTe$_2$ \cite{Kan19.1}, and $T_{\rm d}$-TaIrTe$_4$ \cite{Kum21.1}, \red{as well as carrier-concentration-optimized ferroelectric (Pb$_{1-x}$Sn$_x$)$_{1-y}$In$_y$Te \cite{Zha22.2}.} The quantity that conveniently benchmarks the size of this effect is $\mathcal{E}_{xy}^{2\omega,\rm{intr}}/(\mathcal{E}_x^{\omega})^2$, where $\mathcal{E}_x^{\omega}$ is the applied electric field and $\mathcal{E}_{xy}^{2\omega,\rm{intr}}$ the intrinsic part of the resulting transverse electric field \red{at double frequency.} A comparison of all available data reveals that $\mathcal{E}_{xy}^{2\omega,\rm{intr}}/(\mathcal{E}_x^{\omega})^2$ is \red{drastically enhanced by strong correlations. Furthermore, as it also increases with decreasing charge carrier concentration, a strategy for further boosting the intrinsic topological Hall response is to reduce the carrier concentration of strongly correlated Weyl semimetals. We propose gating experiments on thin films as a promising strategy to explore this route.}

\red{An interesting topic for further studies across the correlation spectrum are nonlinear optical responses, as seen in several noninteracting/weakly interacting Weyl semimetals and discussed also in terms of their potential for applications \cite{Kum21.1,Wu16.1,Wan22.2}. Strongly correlated Weyl semimetals might amplify such responses and reduce the pertinent energies, thereby enabling e.g.\ non-reciprocal devices and rectification in the microwave regime.}

We hope that our comparison of the key characteristics of the Weyl--Kondo semimetal Ce$_3$Bi$_4$Pd$_3$ with features of other candidate materials provides valuable guidance to discover new strongly correlated \red{Weyl semimetals. This would allow the determination of universal aspects in Weyl--Kondo semimetals, such as the dependence of the magnitude of the nonlinear Hall response with the Weyl velocity (correlation strengths), charge carrier concentration (distance of the nodes from the Fermi energy), and potentially other factors such as the node vs anti-node separation in momentum space, tilting, and multiplicity of the Weyl nodes.} This, in turn, may motivate further theoretical development and, more generally, boost progress toward a broader understanding of correlation-driven topological semimetals across different materials classes, \red{as well as the development of technological applications.}

\section*{Acknowledgements}
We thank J.\ Cano, J.\ Checkelsky, G.\ Eguchi, S. Grefe, A.\ Prokofiev, Q.\ Si, X.\ Yan, and D.\ Zocco for fruitful discussions, which were in part conducted at the Kavli Institute for Theoretical Physics at UC Santa Barbara. This work was supported by the Austrian Science Fund (I5868 - FOR 5249 QUAST, F86 - SFB Q-M\&S), the European Union's Horizon 2020 Research and Innovation Programme (824109, EMP), and the European Research Council (ERC Advanced Grant 101055088, CorMeTop), and in part by the US National Science Foundation (Grant No. NSF PHY-1748958). \red{For the purpose of open access, the authors have applied a CC BY public copyright licence to the Author Accepted Manuscript version arising from this submission.}\\


\begin{thebibliography}{10}
\expandafter\ifx\csname url\endcsname\relax
  \def\url#1{\texttt{#1}}\fi
\expandafter\ifx\csname urlprefix\endcsname\relax\def\urlprefix{URL }\fi
\providecommand{\bibinfo}[2]{#2}
\providecommand{\eprint}[2][]{\url{#2}}

\bibitem{Ste84.1}
\bibinfo{author}{Stewart, G.~R.}
\newblock \bibinfo{title}{{Heavy-fermion systems}}.
\newblock \emph{\bibinfo{journal}{{Rev.\ Mod.\ Phys.}}}
  \textbf{\bibinfo{volume}{56}}, \bibinfo{pages}{{755}} (\bibinfo{year}{1984}).

\bibitem{Hew97.1}
\bibinfo{author}{Hewson, A.~C.}
\newblock \emph{\bibinfo{title}{{The Kondo Problem to Heavy Fermions}}}
  (\bibinfo{publisher}{Cambridge University Press},
  \bibinfo{address}{Cambridge}, \bibinfo{year}{1997}).

\bibitem{Loe07.1}
\bibinfo{author}{{v. L\"ohneysen}, H.}, \bibinfo{author}{Rosch, A.},
  \bibinfo{author}{Vojta, M.} \& \bibinfo{author}{W\"olfle, P.}
\newblock \bibinfo{title}{{Fermi-liquid instabilities at magnetic quantum
  critical points}}.
\newblock \emph{\bibinfo{journal}{{Rev.\ Mod.\ Phys.}}}
  \textbf{\bibinfo{volume}{79}}, \bibinfo{pages}{1015} (\bibinfo{year}{2007}).

\bibitem{Kad86.1}
\bibinfo{author}{Kadowaki, K.} \& \bibinfo{author}{Woods, S.~B.}
\newblock \bibinfo{title}{{Universal relationship of the resistivity and
  specific heat in heavy-fermion compounds}}.
\newblock \emph{\bibinfo{journal}{{Solid State Commun.}}}
  \textbf{\bibinfo{volume}{58}}, \bibinfo{pages}{507--509}
  (\bibinfo{year}{1986}).

\bibitem{Jac09.1}
\bibinfo{author}{Jacko, A.~C.}, \bibinfo{author}{Fjaerestad, J.~O.} \&
  \bibinfo{author}{Powell, B.~J.}
\newblock \bibinfo{title}{{A unified explanation of the Kadowaki--Woods ratio
  in strongly correlated metals}}.
\newblock \emph{\bibinfo{journal}{{Nat.\ Phys.}}} \textbf{\bibinfo{volume}{5}},
  \bibinfo{pages}{422} (\bibinfo{year}{2009}).

\bibitem{Sch99.1}
\bibinfo{author}{Schofield, A.}
\newblock \bibinfo{title}{{Non-Fermi liquids}}.
\newblock \emph{\bibinfo{journal}{{Contemp.\ Phys.}}}
  \textbf{\bibinfo{volume}{40}}, \bibinfo{pages}{95} (\bibinfo{year}{1999}).

\bibitem{Col01.1}
\bibinfo{author}{Coleman, P.}, \bibinfo{author}{P\'epin, C.},
  \bibinfo{author}{Si, Q.} \& \bibinfo{author}{Ramazashvili, R.}
\newblock \bibinfo{title}{{How do Fermi liquids get heavy and die?}}
\newblock \emph{\bibinfo{journal}{{J.\ Phys.: Condens.\ Matter}}}
  \textbf{\bibinfo{volume}{13}}, \bibinfo{pages}{R723--R738}
  (\bibinfo{year}{2001}).

\bibitem{Ste01.1}
\bibinfo{author}{Stewart, G.~R.}
\newblock \bibinfo{title}{{Non-Fermi-liquid behavior in $d$- and $f$-electron
  metals}}.
\newblock \emph{\bibinfo{journal}{{Rev.\ Mod.\ Phys.}}}
  \textbf{\bibinfo{volume}{73}}, \bibinfo{pages}{{797}} (\bibinfo{year}{2001}).

\bibitem{Kir20.1}
\bibinfo{author}{Kirchner, S.}, \bibinfo{author}{Paschen, S.},
  \bibinfo{author}{Chen, Q.}, \bibinfo{author}{Wirth, S.},
  \bibinfo{author}{Feng, D.}, \bibinfo{author}{Thompson, J.~D.} \&
  \bibinfo{author}{Si, Q.}
\newblock \bibinfo{title}{Colloquium: Heavy-electron quantum criticality and
  single-particle spectroscopy}.
\newblock \emph{\bibinfo{journal}{Rev. Mod. Phys.}}
  \textbf{\bibinfo{volume}{92}}, \bibinfo{pages}{011002}
  (\bibinfo{year}{2020}).

\bibitem{Pas21.1}
\bibinfo{author}{Paschen, S.} \& \bibinfo{author}{Si, Q.}
\newblock \bibinfo{title}{{Quantum phases driven by strong correlations}}.
\newblock \emph{\bibinfo{journal}{{Nat.\ Rev.\ Phys.}}}
  \textbf{\bibinfo{volume}{3}}, \bibinfo{pages}{9--26} (\bibinfo{year}{2021}).

\bibitem{All78.1}
\bibinfo{author}{Allen, J.~W.}, \bibinfo{author}{Martin, R.~M.},
  \bibinfo{author}{Parc, X.}, \bibinfo{author}{Batlogg, B.} \&
  \bibinfo{author}{Wachter, P.}
\newblock \bibinfo{title}{{Mixed valent SmB$_6$ and gold-SmS: Metals or
  insulators?}}
\newblock \emph{\bibinfo{journal}{{J.\ Appl.\ Phys.}}}
  \textbf{\bibinfo{volume}{49}}, \bibinfo{pages}{2078} (\bibinfo{year}{1978}).

\bibitem{Tra84.1}
\bibinfo{author}{Travaglini, G.} \& \bibinfo{author}{Wachter, P.}
\newblock \bibinfo{title}{{Intermediate-valent SmB$_6$ and the hybridization
  model: An optical study}}.
\newblock \emph{\bibinfo{journal}{{Phys.\ Rev.\ B}}}
  \textbf{\bibinfo{volume}{29}}, \bibinfo{pages}{893} (\bibinfo{year}{1984}).

\bibitem{Aep92.1}
\bibinfo{author}{Aeppli, G.} \& \bibinfo{author}{Fisk, Z.}
\newblock \bibinfo{title}{{Kondo insulators}}.
\newblock \emph{\bibinfo{journal}{{Comments Condens.\ Matter Phys.}}}
  \textbf{\bibinfo{volume}{16}}, \bibinfo{pages}{155} (\bibinfo{year}{1992}).

\bibitem{Buc94.1}
\bibinfo{author}{Bucher, B.}, \bibinfo{author}{Schlesinger, Z.},
  \bibinfo{author}{Canfield, P.~C.} \& \bibinfo{author}{Fisk, Z.}
\newblock \bibinfo{title}{{Kondo coupling induced charge gap in
  Ce$_3$Bi$_4$Pt$_3$}}.
\newblock \emph{\bibinfo{journal}{{Phys.\ Rev.\ Lett.}}}
  \textbf{\bibinfo{volume}{72}}, \bibinfo{pages}{522} (\bibinfo{year}{1994}).

\bibitem{Roz96.1}
\bibinfo{author}{Rozenberg, M.~J.}, \bibinfo{author}{Kotliar, G.} \&
  \bibinfo{author}{Kajueter, H.}
\newblock \bibinfo{title}{{Transfer of spectral weight in spectroscopies of
  correlated electron systems}}.
\newblock \emph{\bibinfo{journal}{Phys. Rev. B}} \textbf{\bibinfo{volume}{54}},
  \bibinfo{pages}{8452--8468} (\bibinfo{year}{1996}).

\bibitem{Tsu97.1}
\bibinfo{author}{Tsunetsugu, H.}, \bibinfo{author}{Sigrist, M.} \&
  \bibinfo{author}{Ueda, K.}
\newblock \bibinfo{title}{{The ground-state phase diagram of the
  one-dimensional Kondo lattice model}}.
\newblock \emph{\bibinfo{journal}{{Rev.\ Mod.\ Phys.}}}
  \textbf{\bibinfo{volume}{69}}, \bibinfo{pages}{809} (\bibinfo{year}{1997}).

\bibitem{Ris00.1}
\bibinfo{author}{Riseborough, P.~S.}
\newblock \bibinfo{title}{{Heavy fermion semiconductors}}.
\newblock \emph{\bibinfo{journal}{{Adv.\ Phys.}}}
  \textbf{\bibinfo{volume}{49}}, \bibinfo{pages}{257} (\bibinfo{year}{2000}).

\bibitem{Yam10.1}
\bibinfo{author}{Yamamoto, S.~J.} \& \bibinfo{author}{Si, Q.}
\newblock \bibinfo{title}{{Global phase diagram of the Kondo lattice: From
  heavy fermion metals to Kondo insulators}}.
\newblock \emph{\bibinfo{journal}{{J.\ Low Temp.\ Phys.}}}
  \textbf{\bibinfo{volume}{{161}}}, \bibinfo{pages}{233}
  (\bibinfo{year}{{2010}}).

\bibitem{Si13.1}
\bibinfo{author}{Si, Q.} \& \bibinfo{author}{Paschen, S.}
\newblock \bibinfo{title}{{Quantum phase transitions in heavy fermion metals
  and Kondo insulators}}.
\newblock \emph{\bibinfo{journal}{{Phys.\ Status Solidi B}}}
  \textbf{\bibinfo{volume}{250}}, \bibinfo{pages}{425} (\bibinfo{year}{2013}).

\bibitem{Ono91.1}
\bibinfo{author}{{\={O}}no, Y.}, \bibinfo{author}{Matsuura, T.} \&
  \bibinfo{author}{Kuroda, Y.}
\newblock \bibinfo{title}{{Electronic state of the Anderson lattice over the
  whole temperature range}}.
\newblock \emph{\bibinfo{journal}{{J.\ Phys.\ Soc.\ Jpn.}}}
  \textbf{\bibinfo{volume}{60}}, \bibinfo{pages}{3475--3500}
  (\bibinfo{year}{1991}).

\bibitem{Pru00.1}
\bibinfo{author}{Pruschke, T.}, \bibinfo{author}{Bulla, R.} \&
  \bibinfo{author}{Jarrell, M.}
\newblock \bibinfo{title}{{Low-energy scale of the periodic Anderson model}}.
\newblock \emph{\bibinfo{journal}{Phys. Rev. B}} \textbf{\bibinfo{volume}{61}},
  \bibinfo{pages}{12799--12809} (\bibinfo{year}{2000}).

\bibitem{Has10.1}
\bibinfo{author}{Hasan, M.~Z.} \& \bibinfo{author}{Kane, C.~L.}
\newblock \bibinfo{title}{{Colloquium: Topological insulators}}.
\newblock \emph{\bibinfo{journal}{Rev. Mod. Phys.}}
  \textbf{\bibinfo{volume}{82}}, \bibinfo{pages}{3045} (\bibinfo{year}{2010}).

\bibitem{Dze10.1}
\bibinfo{author}{Dzero, M.}, \bibinfo{author}{Sun, K.},
  \bibinfo{author}{Galitski, V.} \& \bibinfo{author}{Coleman, P.}
\newblock \bibinfo{title}{{Topological Kondo insulators}}.
\newblock \emph{\bibinfo{journal}{{Phys.\ Rev.\ Lett.}}}
  \textbf{\bibinfo{volume}{104}}, \bibinfo{pages}{106408}
  (\bibinfo{year}{2010}).

\bibitem{Jia13.1}
\bibinfo{author}{Jiang, J.}, \bibinfo{author}{Li, S.}, \bibinfo{author}{Zhang,
  T.}, \bibinfo{author}{Sun, Z.}, \bibinfo{author}{Chen, F.},
  \bibinfo{author}{Ye, Z.~R.}, \bibinfo{author}{Xu, M.}, \bibinfo{author}{Ge,
  Q.~Q.}, \bibinfo{author}{Tan, S.~Y.}, \bibinfo{author}{Niu, X.~H.},
  \bibinfo{author}{Xia, M.}, \bibinfo{author}{Xie, B.~P.}, \bibinfo{author}{Li,
  Y.~F.}, \bibinfo{author}{Chen, X.~H.}, \bibinfo{author}{Wen, H.~H.} \&
  \bibinfo{author}{Feng, D.~L.}
\newblock \bibinfo{title}{{Observation of possible topological in-gap surface
  states in the Kondo insulator SmB$_6$ by photoemission}}.
\newblock \emph{\bibinfo{journal}{Nat.\ Commun.}} \textbf{\bibinfo{volume}{4}},
  \bibinfo{pages}{3010} (\bibinfo{year}{2013}).

\bibitem{Neu13.1}
\bibinfo{author}{Neupane, M.}, \bibinfo{author}{Alidoust, N.},
  \bibinfo{author}{Xu, S.~Y.}, \bibinfo{author}{Kondo, T.},
  \bibinfo{author}{Ishida, Y.}, \bibinfo{author}{Kim, D.~J.},
  \bibinfo{author}{Liu, C.}, \bibinfo{author}{Belopolski, I.},
  \bibinfo{author}{Jo, Y.~J.}, \bibinfo{author}{Chang, T.~R.},
  \bibinfo{author}{Jeng, H.~T.}, \bibinfo{author}{Durakiewicz, T.},
  \bibinfo{author}{Balicas, L.}, \bibinfo{author}{Lin, H.},
  \bibinfo{author}{Bansil, A.}, \bibinfo{author}{Shin, S.},
  \bibinfo{author}{Fisk, Z.} \& \bibinfo{author}{Hasan, M.~Z.}
\newblock \bibinfo{title}{{Surface electronic structure of the topological
  Kondo-insulator candidate correlated electron system SmB$_6$}}.
\newblock \emph{\bibinfo{journal}{Nat.\ Commun.}} \textbf{\bibinfo{volume}{4}},
  \bibinfo{pages}{2991} (\bibinfo{year}{2013}).

\bibitem{Wen14.1}
\bibinfo{author}{Weng, H.}, \bibinfo{author}{Zhao, J.}, \bibinfo{author}{Wang,
  Z.}, \bibinfo{author}{Fang, Z.} \& \bibinfo{author}{Dai, X.}
\newblock \bibinfo{title}{{Topological crystalline Kondo insulator in mixed
  valence ytterbium borides}}.
\newblock \emph{\bibinfo{journal}{Phys.\ Rev.\ Lett.}}
  \textbf{\bibinfo{volume}{112}}, \bibinfo{pages}{016403}
  (\bibinfo{year}{2014}).

\bibitem{Kim14.2}
\bibinfo{author}{Kim, D.~J.}, \bibinfo{author}{Xia, J.} \&
  \bibinfo{author}{Fisk, Z.}
\newblock \bibinfo{title}{{Topological surface state in the Kondo insulator
  samarium hexaboride}}.
\newblock \emph{\bibinfo{journal}{Nature Mat.}} \textbf{\bibinfo{volume}{13}},
  \bibinfo{pages}{466} (\bibinfo{year}{2014}).

\bibitem{Li14.1}
\bibinfo{author}{Li, G.}, \bibinfo{author}{Xiang, Z.}, \bibinfo{author}{Yu,
  F.}, \bibinfo{author}{Asaba, T.}, \bibinfo{author}{Lawson, B.},
  \bibinfo{author}{Cai, P.}, \bibinfo{author}{Tinsman, C.},
  \bibinfo{author}{Berkley, A.}, \bibinfo{author}{Wolgast, S.},
  \bibinfo{author}{Eo, Y.~S.}, \bibinfo{author}{Kim, D.-J.},
  \bibinfo{author}{Kurdak, C.}, \bibinfo{author}{Allen, J.~W.},
  \bibinfo{author}{Sun, K.}, \bibinfo{author}{Chen, X.~H.},
  \bibinfo{author}{Wang, Y.~Y.}, \bibinfo{author}{Fisk, Z.} \&
  \bibinfo{author}{Li, L.}
\newblock \bibinfo{title}{{Two-dimensional Fermi surfaces in Kondo insulator
  SmB$_6$}}.
\newblock \emph{\bibinfo{journal}{Science}} \textbf{\bibinfo{volume}{346}},
  \bibinfo{pages}{1208} (\bibinfo{year}{2014}).

\bibitem{Xu14.1}
\bibinfo{author}{{Xu, N. and Biswas, P. K. and Dil, J. H. and Dhaka, R. S. and
  Landolt, G. and Muff, S. and Matt, C. E. and Shi, X. and Plumb, N. C. and
  Radovic, M. and Pomjakushina, E. and Conder, K. and Amato, A. and Borisenko,
  S. V. and Yu, R. and Weng, H. M. and Fang, Z. and Dai, X. and Mesot, J. and
  Ding, H. and Shi, M.}}
\newblock \bibinfo{title}{{Direct observation of the spin texture in SmB$_6$ as
  evidence of the topological Kondo insulator}}.
\newblock \emph{\bibinfo{journal}{{Nat. Commun.}}}
  \textbf{\bibinfo{volume}{5}}, \bibinfo{pages}{{4566}} (\bibinfo{year}{2014}).

\bibitem{Tan15.1}
\bibinfo{author}{Tan, B.~S.}, \bibinfo{author}{Hsu, Y.-T.},
  \bibinfo{author}{Zeng, B.}, \bibinfo{author}{Hatnean, M.~C.},
  \bibinfo{author}{Harrison, N.}, \bibinfo{author}{Zhu, Z.},
  \bibinfo{author}{Hartstein, M.}, \bibinfo{author}{Kiourlappou, M.},
  \bibinfo{author}{Srivastava, A.}, \bibinfo{author}{Johannes, M.~D.},
  \bibinfo{author}{Murphy, T.~P.}, \bibinfo{author}{Park, J.-H.},
  \bibinfo{author}{Balicas, L.}, \bibinfo{author}{Lonzarich, G.~G.},
  \bibinfo{author}{Balakrishnan, G.} \& \bibinfo{author}{Sebastian, S.~E.}
\newblock \bibinfo{title}{{Unconventional Fermi surface in an insulating
  state}}.
\newblock \emph{\bibinfo{journal}{Science}} \textbf{\bibinfo{volume}{349}},
  \bibinfo{pages}{287} (\bibinfo{year}{2015}).

\bibitem{Dze16.1}
\bibinfo{author}{Dzero, M.}, \bibinfo{author}{Xia, J.},
  \bibinfo{author}{Galitski, V.} \& \bibinfo{author}{Coleman, P.}
\newblock \bibinfo{title}{{Topological Kondo insulators}}.
\newblock \emph{\bibinfo{journal}{{Annu.\ Rev.\ Condens.\ Matter Phys.}}}
  \textbf{\bibinfo{volume}{7}}, \bibinfo{pages}{249} (\bibinfo{year}{2016}).

\bibitem{Nak16.1}
\bibinfo{author}{Nakajima, Y.}, \bibinfo{author}{Syers, P.},
  \bibinfo{author}{Wang, X.}, \bibinfo{author}{Wang, R.} \&
  \bibinfo{author}{Paglione, J.}
\newblock \bibinfo{title}{{One-dimensional edge state transport in a
  topological Kondo insulator}}.
\newblock \emph{\bibinfo{journal}{{Nat. Phys.}}} \textbf{\bibinfo{volume}{12}},
  \bibinfo{pages}{213} (\bibinfo{year}{2016}).

\bibitem{Par16.1}
\bibinfo{author}{Park, W.~K.}, \bibinfo{author}{Sun, L.},
  \bibinfo{author}{Noddings, A.}, \bibinfo{author}{Kim, D.-J.},
  \bibinfo{author}{Fisk, Z.} \& \bibinfo{author}{Greene, L.~H.}
\newblock \bibinfo{title}{{Topological surface states interacting with bulk
  excitations in the Kondo insulator SmB$_6$ revealed via planar tunneling
  spectroscopy}}.
\newblock \emph{\bibinfo{journal}{{Proc.\ Natl.\ Acad.\ Sci.\ U.S.A.}}}
  \textbf{\bibinfo{volume}{113}}, \bibinfo{pages}{6599} (\bibinfo{year}{2016}).

\bibitem{Rac18.1}
\bibinfo{author}{Rachel, S.}
\newblock \bibinfo{title}{{Interacting topological insulators: a review}}.
\newblock \emph{\bibinfo{journal}{{Rep.\ Prog.\ Phys.}}}
  \textbf{\bibinfo{volume}{81}}, \bibinfo{pages}{116501}
  (\bibinfo{year}{2018}).

\bibitem{Pir20.1}
\bibinfo{author}{Pirie, H.}, \bibinfo{author}{Liu, Y.},
  \bibinfo{author}{Soumyanarayanan, A.}, \bibinfo{author}{Chen, P.},
  \bibinfo{author}{He, Y.}, \bibinfo{author}{Yee, M.~M.},
  \bibinfo{author}{Rosa, P. F.~S.}, \bibinfo{author}{Thompson, J.~D.},
  \bibinfo{author}{Kim, D.-J.}, \bibinfo{author}{Fisk, Z.},
  \bibinfo{author}{Wang, X.}, \bibinfo{author}{Paglione, J.},
  \bibinfo{author}{Morr, D.~K.}, \bibinfo{author}{Hamidian, M.~H.} \&
  \bibinfo{author}{Hoffman, J.~E.}
\newblock \bibinfo{title}{{Imaging emergent heavy Dirac fermions of a
  topological Kondo insulator}}.
\newblock \emph{\bibinfo{journal}{{Nat.\ Phys.}}}
  \textbf{\bibinfo{volume}{{16}}}, \bibinfo{pages}{{52}}
  (\bibinfo{year}{{2020}}).

\bibitem{Li20.2}
\bibinfo{author}{Li, L.}, \bibinfo{author}{Sun, K.}, \bibinfo{author}{Kurdak,
  C.} \& \bibinfo{author}{Allen, J.~W.}
\newblock \bibinfo{title}{{Emergent mystery in the Kondo insulator samarium
  hexaboride}}.
\newblock \emph{\bibinfo{journal}{{Nat.\ Rev.\ Phys.}}}
  \textbf{\bibinfo{volume}{2}}, \bibinfo{pages}{463--479}
  (\bibinfo{year}{2020}).

\bibitem{Ais22.1}
\bibinfo{author}{Aishwarya, A.}, \bibinfo{author}{Cai, Z.},
  \bibinfo{author}{Raghavan, A.}, \bibinfo{author}{Romanelli, M.},
  \bibinfo{author}{Wang, X.}, \bibinfo{author}{Li, X.}, \bibinfo{author}{Gu,
  G.~D.}, \bibinfo{author}{Hirsbrunner, M.}, \bibinfo{author}{Hughes, T.},
  \bibinfo{author}{Liu, F.}, \bibinfo{author}{Jiao, L.} \&
  \bibinfo{author}{Madhavan, V.}
\newblock \bibinfo{title}{{Spin-selective tunneling from nanowires of the
  candidate topological Kondo insulator SmB$_6$}}.
\newblock \emph{\bibinfo{journal}{Science}} \textbf{\bibinfo{volume}{377}},
  \bibinfo{pages}{1218--1222} (\bibinfo{year}{2022}).

\bibitem{Dzs17.1}
\bibinfo{author}{Dzsaber, S.}, \bibinfo{author}{Prochaska, L.},
  \bibinfo{author}{Sidorenko, A.}, \bibinfo{author}{Eguchi, G.},
  \bibinfo{author}{Svagera, R.}, \bibinfo{author}{Waas, M.},
  \bibinfo{author}{Prokofiev, A.}, \bibinfo{author}{Si, Q.} \&
  \bibinfo{author}{Paschen, S.}
\newblock \bibinfo{title}{{Kondo insulator to semimetal transformation tuned by
  spin-orbit coupling}}.
\newblock \emph{\bibinfo{journal}{{Phys.\ Rev.\ Lett.}}}
  \textbf{\bibinfo{volume}{118}}, \bibinfo{pages}{246601}
  (\bibinfo{year}{2017}).

\bibitem{Lai18.1}
\bibinfo{author}{Lai, H.-H.}, \bibinfo{author}{Grefe, S.~E.},
  \bibinfo{author}{Paschen, S.} \& \bibinfo{author}{Si, Q.}
\newblock \bibinfo{title}{{Weyl-Kondo semimetal in heavy-fermion systems}}.
\newblock \emph{\bibinfo{journal}{{Proc.\ Natl.\ Acad.\ Sci.\ U.S.A.}}}
  \textbf{\bibinfo{volume}{115}}, \bibinfo{pages}{93} (\bibinfo{year}{2018}).

\bibitem{Gre20.1}
\bibinfo{author}{Grefe, S.~E.}, \bibinfo{author}{Lai, H.-H.},
  \bibinfo{author}{Paschen, S.} \& \bibinfo{author}{Si, Q.}
\newblock \bibinfo{title}{{Weyl-Kondo semimetals in nonsymmorphic systems}}.
\newblock \emph{\bibinfo{journal}{Phys. Rev. B}}
  \textbf{\bibinfo{volume}{101}}, \bibinfo{pages}{075138}
  (\bibinfo{year}{2020}).

\bibitem{Dzs21.1}
\bibinfo{author}{Dzsaber, S.}, \bibinfo{author}{Yan, X.},
  \bibinfo{author}{Eguchi, G.}, \bibinfo{author}{Prokofiev, A.},
  \bibinfo{author}{Shiroka, T.}, \bibinfo{author}{Blaha, P.},
  \bibinfo{author}{Rubel, O.}, \bibinfo{author}{Grefe, S.~E.},
  \bibinfo{author}{Lai, H.-H.}, \bibinfo{author}{Si, Q.} \&
  \bibinfo{author}{Paschen, S.}
\newblock \bibinfo{title}{{Giant spontaneous Hall effect in a nonmagnetic
  Weyl-Kondo semimetal}}.
\newblock \emph{\bibinfo{journal}{{Proc. Natl. Acad. Sci. U.S.A.}}}
  \textbf{\bibinfo{volume}{118}}, \bibinfo{pages}{e2013386118}
  (\bibinfo{year}{2021}).

\bibitem{Che22.1}
\bibinfo{author}{Chen, L.}, \bibinfo{author}{Setty, C.}, \bibinfo{author}{Hu,
  H.}, \bibinfo{author}{Vergniory, M.~G.}, \bibinfo{author}{Grefe, S.~E.},
  \bibinfo{author}{Fischer, L.}, \bibinfo{author}{Yan, X.},
  \bibinfo{author}{Eguchi, G.}, \bibinfo{author}{Prokofiev, A.},
  \bibinfo{author}{Paschen, S.}, \bibinfo{author}{Cano, J.} \&
  \bibinfo{author}{Si, Q.}
\newblock \bibinfo{title}{{Topological semimetal driven by strong correlations
  and crystalline symmetry}}.
\newblock \emph{\bibinfo{journal}{{Nat.\ Phys.}}}
  \textbf{\bibinfo{volume}{18}}, \bibinfo{pages}{1341--1346}
  (\bibinfo{year}{2022}).

\bibitem{Arm18.1}
\bibinfo{author}{Armitage, N.~P.}, \bibinfo{author}{Mele, E.~J.} \&
  \bibinfo{author}{Vishwanath, A.}
\newblock \bibinfo{title}{{Weyl and Dirac semimetals in three-dimensional
  solids}}.
\newblock \emph{\bibinfo{journal}{{Rev.\ Mod.\ Phys.}}}
  \textbf{\bibinfo{volume}{90}}, \bibinfo{pages}{015001}
  (\bibinfo{year}{2018}).

\bibitem{Has21.1}
\bibinfo{author}{Hasan, M.~Z.}, \bibinfo{author}{Chang, G.},
  \bibinfo{author}{Belopolski, I.}, \bibinfo{author}{Bian, G.},
  \bibinfo{author}{Xu, S.-Y.} \& \bibinfo{author}{Yin, J.-X.}
\newblock \bibinfo{title}{{Weyl, Dirac and high-fold chiral fermions in
  topological quantum matter}}.
\newblock \emph{\bibinfo{journal}{{Nat.\ Rev.\ Mater.}}}
  \textbf{\bibinfo{volume}{6}}, \bibinfo{pages}{784--803}
  (\bibinfo{year}{2021}).

\bibitem{Dzs22.1}
\bibinfo{author}{Dzsaber, S.}, \bibinfo{author}{Zocco, D.~A.},
  \bibinfo{author}{McCollam, A.}, \bibinfo{author}{Weickert, F.},
  \bibinfo{author}{McDonald, R.}, \bibinfo{author}{Taupin, M.},
  \bibinfo{author}{Yan, X.}, \bibinfo{author}{Prokofiev, A.},
  \bibinfo{author}{Tang, L. M.~K.}, \bibinfo{author}{Vlaar, B.},
  \bibinfo{author}{Winter, L.~E.}, \bibinfo{author}{Jaime, M.},
  \bibinfo{author}{Si, Q.} \& \bibinfo{author}{Paschen, S.}
\newblock \bibinfo{title}{{Control of electronic topology in a strongly
  correlated electron system}}.
\newblock \emph{\bibinfo{journal}{{Nat.\ Commun.}}}
  \textbf{\bibinfo{volume}{13}}, \bibinfo{pages}{5729} (\bibinfo{year}{2022}).

\bibitem{Gre20.1x}
\bibinfo{author}{Grefe, S.~E.}, \bibinfo{author}{Lai, H.-H.},
  \bibinfo{author}{Paschen, S.} \& \bibinfo{author}{Si, Q.}
\newblock \bibinfo{title}{{Extreme response of Weyl-Kondo semimetal to Zeeman
  coupling. {\em arXiv:2012.15841} (2020)}}.

\bibitem{Hun90.1}
\bibinfo{author}{Hundley, M.~F.}, \bibinfo{author}{Canfield, P.~C.},
  \bibinfo{author}{Thompson, J.~D.}, \bibinfo{author}{Fisk, Z.} \&
  \bibinfo{author}{Lawrence, J.~M.}
\newblock \bibinfo{title}{{Hybridization gap in Ce$_3$Bi$_4$Pt$_3$}}.
\newblock \emph{\bibinfo{journal}{{Phys.\ Rev.\ B}}}
  \textbf{\bibinfo{volume}{42}}, \bibinfo{pages}{6842} (\bibinfo{year}{1990}).

\bibitem{Rey94.1}
\bibinfo{author}{Reyes, A.~P.}, \bibinfo{author}{Heffner, R.~H.},
  \bibinfo{author}{Canfield, P.~C.}, \bibinfo{author}{Thompson, J.~D.} \&
  \bibinfo{author}{Fisk, Z.}
\newblock \bibinfo{title}{{$^{209}$Bi NMR and NQR investigation of the
  small-gap semiconductor Ce$_3$Bi$_4$Pt$_3$}}.
\newblock \emph{\bibinfo{journal}{{Phys.\ Rev.\ B}}}
  \textbf{\bibinfo{volume}{49}}, \bibinfo{pages}{16321} (\bibinfo{year}{1994}).

\bibitem{Jai00.1}
\bibinfo{author}{Jaime, M.}, \bibinfo{author}{Movshovich, R.},
  \bibinfo{author}{Stewart, G.}, \bibinfo{author}{Beyermann, W.},
  \bibinfo{author}{Berisso, M.}, \bibinfo{author}{Hundley, M.},
  \bibinfo{author}{Canfield, P.} \& \bibinfo{author}{Sarrao, J.}
\newblock \bibinfo{title}{{Closing the spin gap in the Kondo insulator
  Ce$_3$Bi$_4$Pt$_3$ at high magnetic fields}}.
\newblock \emph{\bibinfo{journal}{{Nature}}} \textbf{\bibinfo{volume}{{405}}},
  \bibinfo{pages}{160} (\bibinfo{year}{{2000}}).

\bibitem{Kus19.1}
\bibinfo{author}{Kushwaha, S.~K.}, \bibinfo{author}{Chan, M.~K.},
  \bibinfo{author}{Park, J.}, \bibinfo{author}{Thomas, S.~M.},
  \bibinfo{author}{Bauer, E.~D.}, \bibinfo{author}{Thompson, J.~D.},
  \bibinfo{author}{Ronning, F.}, \bibinfo{author}{Rosa, P. F.~S.} \&
  \bibinfo{author}{Harrison, N.}
\newblock \bibinfo{title}{{Magnetic field-tuned Fermi liquid in a Kondo
  insulator}}.
\newblock \emph{\bibinfo{journal}{{Nat.\ Commun.}}}
  \textbf{\bibinfo{volume}{10}}, \bibinfo{pages}{5487} (\bibinfo{year}{2019}).

\bibitem{Sod15.1}
\bibinfo{author}{Sodemann, I.} \& \bibinfo{author}{Fu, L.}
\newblock \bibinfo{title}{{Quantum nonlinear Hall effect induced by Berry
  curvature dipole in time-reversal invariant materials}}.
\newblock \emph{\bibinfo{journal}{{Phys.\ Rev.\ Lett.}}}
  \textbf{\bibinfo{volume}{115}}, \bibinfo{pages}{216806}
  (\bibinfo{year}{2015}).

\bibitem{Cas09.1}
\bibinfo{author}{Castro~Neto, A.~H.}, \bibinfo{author}{Guinea, F.},
  \bibinfo{author}{Peres, N. M.~R.}, \bibinfo{author}{Novoselov, K.~S.} \&
  \bibinfo{author}{Geim, A.~K.}
\newblock \bibinfo{title}{{The electronic properties of graphene}}.
\newblock \emph{\bibinfo{journal}{{Rev.\ Mod.\ Phys.}}}
  \textbf{\bibinfo{volume}{81}}, \bibinfo{pages}{109--162}
  (\bibinfo{year}{2009}).

\bibitem{Guo18.1}
\bibinfo{author}{Guo, C.~Y.}, \bibinfo{author}{Wu, F.}, \bibinfo{author}{Wu,
  Z.~Z.}, \bibinfo{author}{Smidman, M.}, \bibinfo{author}{Cao, C.},
  \bibinfo{author}{Bostwick, A.}, \bibinfo{author}{Jozwiak, C.},
  \bibinfo{author}{Rotenberg, E.}, \bibinfo{author}{Liu, Y.},
  \bibinfo{author}{Steglich, F.} \& \bibinfo{author}{Yuan, H.~Q.}
\newblock \bibinfo{title}{{Evidence for Weyl fermions in a canonical
  heavy-fermion semimetal YbPtBi}}.
\newblock \emph{\bibinfo{journal}{{Nat.\ Commun.}}}
  \textbf{\bibinfo{volume}{9}}, \bibinfo{pages}{4622} (\bibinfo{year}{2018}).

\bibitem{Hod18.1}
\bibinfo{author}{Hodovanets, H.}, \bibinfo{author}{Eckberg, C.~J.},
  \bibinfo{author}{Zavalij, P.~Y.}, \bibinfo{author}{Kim, H.},
  \bibinfo{author}{Lin, W.-C.}, \bibinfo{author}{Zic, M.},
  \bibinfo{author}{Campbell, D.~J.}, \bibinfo{author}{Higgins, J.~S.} \&
  \bibinfo{author}{Paglione, J.}
\newblock \bibinfo{title}{{Single-crystal investigation of the proposed type-II
  Weyl semimetal CeAlGe}}.
\newblock \emph{\bibinfo{journal}{Phys. Rev. B}} \textbf{\bibinfo{volume}{98}},
  \bibinfo{pages}{245132} (\bibinfo{year}{2018}).

\bibitem{Sin20.1}
\bibinfo{author}{Singh, K.} \& \bibinfo{author}{Mukherjee, K.}
\newblock \bibinfo{title}{{Spin-lattice relaxation phenomena in the magnetic
  state of a suggested Weyl semimetal CeAlGe}}.
\newblock \emph{\bibinfo{journal}{Philos. Mag.}}
  \textbf{\bibinfo{volume}{100}}, \bibinfo{pages}{1771--1787}
  (\bibinfo{year}{2020}).

\bibitem{Cor21.1}
\bibinfo{author}{Corasaniti, M.}, \bibinfo{author}{Yang, R.},
  \bibinfo{author}{Hu, Z.}, \bibinfo{author}{Abeykoon, M.},
  \bibinfo{author}{Petrovic, C.} \& \bibinfo{author}{Degiorgi, L.}
\newblock \bibinfo{title}{{Evidence for correlation effects in
  noncentrosymmetric type-II Weyl semimetals}}.
\newblock \emph{\bibinfo{journal}{Phys. Rev. B}}
  \textbf{\bibinfo{volume}{104}}, \bibinfo{pages}{L121112}
  (\bibinfo{year}{2021}).

\bibitem{Kuo18.1}
\bibinfo{author}{Kuo, C.~N.}, \bibinfo{author}{Chen, W.~T.},
  \bibinfo{author}{Tseng, C.~W.}, \bibinfo{author}{Hsu, C.~J.},
  \bibinfo{author}{Huang, R.~Y.}, \bibinfo{author}{Chou, F.~C.},
  \bibinfo{author}{Kuo, Y.~K.} \& \bibinfo{author}{Lue, C.~S.}
\newblock \bibinfo{title}{{Evidence for a second-order phase transition around
  350 K in ${\mathrm{Ce}}_{3}{\mathrm{Rh}}_{4}{\mathrm{Sn}}_{13}$}}.
\newblock \emph{\bibinfo{journal}{Phys. Rev. B}} \textbf{\bibinfo{volume}{97}},
  \bibinfo{pages}{094101} (\bibinfo{year}{2018}).

\bibitem{Iwa23.1}
\bibinfo{author}{Iwasa, K.}, \bibinfo{author}{Suyama, K.},
  \bibinfo{author}{Ohira-Kawamura, S.}, \bibinfo{author}{Nakajima, K.},
  \bibinfo{author}{Raymond, S.}, \bibinfo{author}{Steffens, P.},
  \bibinfo{author}{Yamada, A.}, \bibinfo{author}{Matsuda, T.~D.},
  \bibinfo{author}{Aoki, Y.}, \bibinfo{author}{Kawasaki, I.},
  \bibinfo{author}{Fujimori, S.-i.}, \bibinfo{author}{Yamagami, H.} \&
  \bibinfo{author}{Yokoyama, M.}
\newblock \bibinfo{title}{{Weyl--Kondo semimetal behavior in the chiral
  structure phase of ${\mathrm{Ce}}_{3}{\mathrm{Rh}}_{4}{\mathrm{Sn}}_{13}$}}.
\newblock \emph{\bibinfo{journal}{Phys. Rev. Mater.}}
  \textbf{\bibinfo{volume}{7}}, \bibinfo{pages}{014201} (\bibinfo{year}{2023}).

\bibitem{Ber79.1}
\bibinfo{author}{Berton, A.}, \bibinfo{author}{Chaussy, J.},
  \bibinfo{author}{Chouteau, G.}, \bibinfo{author}{Cornut, B.},
  \bibinfo{author}{Flouquet, J.}, \bibinfo{author}{Odin, J.},
  \bibinfo{author}{Palleau, J.}, \bibinfo{author}{Peyrard, J.} \&
  \bibinfo{author}{Tournier, R.}
\newblock \bibinfo{title}{{Magnetization and specific heat of abnormal cerium
  compounds}}.
\newblock \emph{\bibinfo{journal}{{J. Phys. Colloq.}}}
  \textbf{\bibinfo{volume}{40}}, \bibinfo{pages}{C5--326}
  (\bibinfo{year}{1979}).

\bibitem{Cor01.1}
\bibinfo{author}{Cornelius, A.~L.}, \bibinfo{author}{Pagliuso, P.~G.},
  \bibinfo{author}{Hundley, M.~F.} \& \bibinfo{author}{Sarrao, J.~L.}
\newblock \bibinfo{title}{{Field-induced magnetic transitions in the
  quasi-two-dimensional heavy-fermion antiferromagnets
  ${\mathrm{Ce}}_{n}{\mathrm{RhIn}}_{3n+2}$ $(n=1$ or 2)}}.
\newblock \emph{\bibinfo{journal}{Phys. Rev. B}} \textbf{\bibinfo{volume}{64}},
  \bibinfo{pages}{144411} (\bibinfo{year}{2001}).

\bibitem{Moc96.1}
\bibinfo{author}{Mock, S.}, \bibinfo{author}{Pietrus, T.},
  \bibinfo{author}{Sidorenko, A.}, \bibinfo{author}{Vollmer, R.} \&
  \bibinfo{author}{L\"ohneysen, H.~v.}
\newblock \bibinfo{title}{{Low-temperature magnetic and thermal properties of
  CePd$_2$In in magnetic fields}}.
\newblock \emph{\bibinfo{journal}{J. Low Temp. Phys.}}
  \textbf{\bibinfo{volume}{104}}, \bibinfo{pages}{95--107}
  (\bibinfo{year}{1996}).

\bibitem{Bud14.1}
\bibinfo{author}{Bud\char39{}ko, S.~L.}, \bibinfo{author}{Hodovanets, H.},
  \bibinfo{author}{Panchula, A.}, \bibinfo{author}{Prozorov, R.} \&
  \bibinfo{author}{Canfield, P.~C.}
\newblock \bibinfo{title}{{Physical properties of CeGe$_{2-x}$ (x=0.24) single
  crystals}}.
\newblock \emph{\bibinfo{journal}{J. Phys. Condens. Matter}}
  \textbf{\bibinfo{volume}{26}}, \bibinfo{pages}{146005}
  (\bibinfo{year}{2014}).

\bibitem{Deo73.1}
\bibinfo{author}{Deonarine, S.} \& \bibinfo{author}{Joshua, S.~J.}
\newblock \bibinfo{title}{{Magnetic field effects on the spin wave spectra and
  magnon specific heat of antiferromagnetic NiF$_2$}}.
\newblock \emph{\bibinfo{journal}{{Phys.\ Stat.\ Solidi (b)}}}
  \textbf{\bibinfo{volume}{57}}, \bibinfo{pages}{767--772}
  (\bibinfo{year}{1973}).

\bibitem{Map06.1}
\bibinfo{author}{Maple, M.~B.}, \bibinfo{author}{Butch, N.~P.},
  \bibinfo{author}{Frederick, N.~A.}, \bibinfo{author}{Ho, P.-C.},
  \bibinfo{author}{Jeffries, J.~R.}, \bibinfo{author}{Saylesa, T.~A.},
  \bibinfo{author}{Yanagisawa, T.}, \bibinfo{author}{Yuhasz, W.~M.},
  \bibinfo{author}{Chi, S.}, \bibinfo{author}{Kang, H.~J.},
  \bibinfo{author}{Lynn, J.~W.}, \bibinfo{author}{Dai, P.},
  \bibinfo{author}{McCall, S.~K.}, \bibinfo{author}{McElfresh, M.~W.},
  \bibinfo{author}{Fluss, M.~J.}, \bibinfo{author}{Henkie, Z.} \&
  \bibinfo{author}{Pietraszko, A.}
\newblock \bibinfo{title}{{Field-dependent ordered phases and Kondo phenomena
  in the filled skutterudite compound PrOs$_4$As$_{12}$}}.
\newblock \emph{\bibinfo{journal}{{Proc.\ Natl.\ Acad.\ Sci.\ U.S.A.}}}
  \textbf{\bibinfo{volume}{103}}, \bibinfo{pages}{6783} (\bibinfo{year}{2006}).

\bibitem{Pup20.1}
\bibinfo{author}{Puphal, P.}, \bibinfo{author}{Pomjakushin, V.},
  \bibinfo{author}{Kanazawa, N.}, \bibinfo{author}{Ukleev, V.},
  \bibinfo{author}{Gawryluk, D.~J.}, \bibinfo{author}{Ma, J.},
  \bibinfo{author}{Naamneh, M.}, \bibinfo{author}{Plumb, N.~C.},
  \bibinfo{author}{Keller, L.}, \bibinfo{author}{Cubitt, R.},
  \bibinfo{author}{Pomjakushina, E.} \& \bibinfo{author}{White, J.~S.}
\newblock \bibinfo{title}{{Topological Magnetic Phase in the Candidate Weyl
  Semimetal CeAlGe}}.
\newblock \emph{\bibinfo{journal}{Phys. Rev. Lett.}}
  \textbf{\bibinfo{volume}{124}}, \bibinfo{pages}{017202}
  (\bibinfo{year}{2020}).

\bibitem{Odu07.1}
\bibinfo{author}{\={O}duchi, Y.}, \bibinfo{author}{Tonohiro, C.},
  \bibinfo{author}{Thamizhavel, A.}, \bibinfo{author}{Nakashima, H.},
  \bibinfo{author}{Morimoto, S.}, \bibinfo{author}{Matsuda, T.},
  \bibinfo{author}{Haga, Y.}, \bibinfo{author}{Sugiyama, K.},
  \bibinfo{author}{Takeuchi, T.}, \bibinfo{author}{Settai, R.},
  \bibinfo{author}{Hagiwara, M.} \& \bibinfo{author}{\={O}nuki, Y.}
\newblock \bibinfo{title}{{Magnetic properties of Ce$_3$T$_4$Sn$_{13}$ and
  Pr$_3$T$_4$Sn$_{13}$ (T=Co and Rh) single crystals}}.
\newblock \emph{\bibinfo{journal}{{J.\ Magn.\ Magn.\ Mater.}}}
  \textbf{\bibinfo{volume}{310}}, \bibinfo{pages}{249--251}
  (\bibinfo{year}{2007}).

\bibitem{Suy18.1}
\bibinfo{author}{Suyama, K.}, \bibinfo{author}{Iwasa, K.},
  \bibinfo{author}{Otomo, Y.}, \bibinfo{author}{Tomiyasu, K.},
  \bibinfo{author}{Sagayama, H.}, \bibinfo{author}{Sagayama, R.},
  \bibinfo{author}{Nakao, H.}, \bibinfo{author}{Kumai, R.},
  \bibinfo{author}{Kitajima, Y.}, \bibinfo{author}{Damay, F. m.~c.},
  \bibinfo{author}{Mignot, J.-M.}, \bibinfo{author}{Yamada, A.},
  \bibinfo{author}{Matsuda, T.~D.} \& \bibinfo{author}{Aoki, Y.}
\newblock \bibinfo{title}{{Chiral-crystal-structure transformations and
  magnetic states of ${R}_{3}{\mathrm{Rh}}_{4}{\mathrm{Sn}}_{13}$
  ($R=\mathrm{La}$ and Ce)}}.
\newblock \emph{\bibinfo{journal}{Phys. Rev. B}} \textbf{\bibinfo{volume}{97}},
  \bibinfo{pages}{235138} (\bibinfo{year}{2018}).

\bibitem{Mun13.1}
\bibinfo{author}{Mun, E.~D.}, \bibinfo{author}{Bud'ko, S.~L.},
  \bibinfo{author}{Martin, C.}, \bibinfo{author}{Kim, H.},
  \bibinfo{author}{Tanatar, M.~A.}, \bibinfo{author}{Park, J.-H.},
  \bibinfo{author}{Murphy, T.}, \bibinfo{author}{Schmiedeshoff, G.~M.},
  \bibinfo{author}{Dilley, N.}, \bibinfo{author}{Prozorov, R.} \&
  \bibinfo{author}{Canfield, P.~C.}
\newblock \bibinfo{title}{{Magnetic-field-tuned quantum criticality of the
  heavy-fermion system YbPtBi}}.
\newblock \emph{\bibinfo{journal}{Phys. Rev. B}} \textbf{\bibinfo{volume}{87}},
  \bibinfo{pages}{075120} (\bibinfo{year}{2013}).

\bibitem{Tia09.1}
\bibinfo{author}{Tian, Y.}, \bibinfo{author}{Ye, L.} \& \bibinfo{author}{Jin,
  X.}
\newblock \bibinfo{title}{{Proper scaling of the anomalous Hall effect}}.
\newblock \emph{\bibinfo{journal}{Phys. Rev. Lett.}}
  \textbf{\bibinfo{volume}{103}}, \bibinfo{pages}{087206}
  (\bibinfo{year}{2009}).

\bibitem{Sch17.1}
\bibinfo{author}{Schilling, M.~B.}, \bibinfo{author}{L\"ohle, A.},
  \bibinfo{author}{Neubauer, D.}, \bibinfo{author}{Shekhar, C.},
  \bibinfo{author}{Felser, C.}, \bibinfo{author}{Dressel, M.} \&
  \bibinfo{author}{Pronin, A.~V.}
\newblock \bibinfo{title}{{Two-channel conduction in YbPtBi}}.
\newblock \emph{\bibinfo{journal}{Phys. Rev. B}} \textbf{\bibinfo{volume}{95}},
  \bibinfo{pages}{155201} (\bibinfo{year}{2017}).

\bibitem{Koe07.2}
\bibinfo{author}{K\"ohler, U.}, \bibinfo{author}{Pikul, A.~P.},
  \bibinfo{author}{Oeschler, N.}, \bibinfo{author}{Westerkamp, T.},
  \bibinfo{author}{Strydom, A.~M.} \& \bibinfo{author}{Steglich, F.}
\newblock \bibinfo{title}{{Low-temperature study of the strongly correlated
  compound Ce$_3$Rh$_4$Sn$_{13}$}}.
\newblock \emph{\bibinfo{journal}{J.\ Phys.\ Condens.\ Matter}}
  \textbf{\bibinfo{volume}{19}}, \bibinfo{pages}{386207}
  (\bibinfo{year}{2007}).

\bibitem{Kan19.1}
\bibinfo{author}{Kang, K.}, \bibinfo{author}{Li, T.}, \bibinfo{author}{Sohn,
  E.}, \bibinfo{author}{Shan, J.} \& \bibinfo{author}{Mak, K.~F.}
\newblock \bibinfo{title}{{Nonlinear anomalous Hall effect in few-layer
  WTe$_2$}}.
\newblock \emph{\bibinfo{journal}{{Nat.\ Mater.}}}
  \textbf{\bibinfo{volume}{18}}, \bibinfo{pages}{324} (\bibinfo{year}{2019}).

\bibitem{Kum21.1}
\bibinfo{author}{Kumar, D.}, \bibinfo{author}{Hsu, C.-H.},
  \bibinfo{author}{Sharma, R.}, \bibinfo{author}{Chang, T.-R.},
  \bibinfo{author}{Yu, P.}, \bibinfo{author}{Wang, J.}, \bibinfo{author}{Eda,
  G.}, \bibinfo{author}{Liang, G.} \& \bibinfo{author}{Yang, H.}
\newblock \bibinfo{title}{{Room-temperature nonlinear Hall effect and wireless
  radiofrequency rectification in Weyl semimetal TaIrTe$_4$}}.
\newblock \emph{\bibinfo{journal}{Nat. Nanotechnol.}}
  \textbf{\bibinfo{volume}{16}}, \bibinfo{pages}{421--425}
  (\bibinfo{year}{2021}).

\bibitem{Ma19.1}
\bibinfo{author}{Ma, Q.}, \bibinfo{author}{Xu, S.-Y.}, \bibinfo{author}{Shen,
  H.}, \bibinfo{author}{MacNeill, D.}, \bibinfo{author}{Fatemi, V.},
  \bibinfo{author}{Chang, T.-R.}, \bibinfo{author}{Mier~Valdivia, A.~M.},
  \bibinfo{author}{Wu, S.}, \bibinfo{author}{Du, Z.}, \bibinfo{author}{Hsu,
  C.-H.}, \bibinfo{author}{Fang, S.}, \bibinfo{author}{Gibson, Q.~D.},
  \bibinfo{author}{Watanabe, K.}, \bibinfo{author}{Taniguchi, T.},
  \bibinfo{author}{Cava, R.~J.}, \bibinfo{author}{Kaxiras, E.},
  \bibinfo{author}{Lu, H.-Z.}, \bibinfo{author}{Lin, H.}, \bibinfo{author}{Fu,
  L.}, \bibinfo{author}{Gedik, N.} \& \bibinfo{author}{Jarillo-Herrero, P.}
\newblock \bibinfo{title}{{Observation of the nonlinear Hall effect under
  time-reversal-symmetric conditions}}.
\newblock \emph{\bibinfo{journal}{Nature}} \textbf{\bibinfo{volume}{565}},
  \bibinfo{pages}{337} (\bibinfo{year}{2019}).

\bibitem{He21.1}
\bibinfo{author}{He, P.}, \bibinfo{author}{Isobe, H.}, \bibinfo{author}{Zhu,
  D.}, \bibinfo{author}{Hsu, C.-H.}, \bibinfo{author}{Fu, L.} \&
  \bibinfo{author}{Yang, H.}
\newblock \bibinfo{title}{{Quantum frequency doubling in the topological
  insulator Bi$_2$Se$_3$}}.
\newblock \emph{\bibinfo{journal}{Nat. Commun.}} \textbf{\bibinfo{volume}{12}},
  \bibinfo{pages}{698} (\bibinfo{year}{2021}).

\bibitem{He22.1}
\bibinfo{author}{He, P.}, \bibinfo{author}{Koon, G. K.~W.},
  \bibinfo{author}{Isobe, H.}, \bibinfo{author}{Tan, J.~Y.},
  \bibinfo{author}{Hu, J.}, \bibinfo{author}{Neto, A. H.~C.},
  \bibinfo{author}{Fu, L.} \& \bibinfo{author}{Yang, H.}
\newblock \bibinfo{title}{{Graphene moir{\'{e}} superlattices with giant
  quantum nonlinearity of chiral Bloch electrons}}.
\newblock \emph{\bibinfo{journal}{Nat. Nanotechnol.}}
  \textbf{\bibinfo{volume}{17}}, \bibinfo{pages}{378--383}
  (\bibinfo{year}{2022}).

\bibitem{Ho21.1}
\bibinfo{author}{Ho, S.-C.}, \bibinfo{author}{Chang, C.-H.},
  \bibinfo{author}{Hsieh, Y.-C.}, \bibinfo{author}{Lo, S.-T.},
  \bibinfo{author}{Huang, B.}, \bibinfo{author}{Vu, T.-H.-Y.},
  \bibinfo{author}{Ortix, C.} \& \bibinfo{author}{Chen, T.-M.}
\newblock \bibinfo{title}{{Hall effects in artificially corrugated bilayer
  graphene without breaking time-reversal symmetry}}.
\newblock \emph{\bibinfo{journal}{Nat. Electron.}}
  \textbf{\bibinfo{volume}{4}}, \bibinfo{pages}{116--125}
  (\bibinfo{year}{2021}).

\bibitem{Tiw21.1}
\bibinfo{author}{Tiwari, A.}, \bibinfo{author}{Chen, F.},
  \bibinfo{author}{Zhong, S.}, \bibinfo{author}{Drueke, E.},
  \bibinfo{author}{Koo, J.}, \bibinfo{author}{Kaczmarek, A.},
  \bibinfo{author}{Xiao, C.}, \bibinfo{author}{Gao, J.}, \bibinfo{author}{Luo,
  X.}, \bibinfo{author}{Niu, Q.}, \bibinfo{author}{Sun, Y.},
  \bibinfo{author}{Yan, B.}, \bibinfo{author}{Zhao, L.} \&
  \bibinfo{author}{Tsen, A.~W.}
\newblock \bibinfo{title}{{Giant c-axis nonlinear anomalous Hall effect in
  Td-MoTe$_2$ and WTe$_2$}}.
\newblock \emph{\bibinfo{journal}{Nat. Commun.}} \textbf{\bibinfo{volume}{12}},
  \bibinfo{pages}{2049} (\bibinfo{year}{2021}).

\bibitem{Dua22.1}
\bibinfo{author}{Duan, J.}, \bibinfo{author}{Jian, Y.}, \bibinfo{author}{Gao,
  Y.}, \bibinfo{author}{Peng, H.}, \bibinfo{author}{Zhong, J.},
  \bibinfo{author}{Feng, Q.}, \bibinfo{author}{Mao, J.} \&
  \bibinfo{author}{Yao, Y.}
\newblock \bibinfo{title}{{Giant Second-Order Nonlinear Hall Effect in Twisted
  Bilayer Graphene}}.
\newblock \emph{\bibinfo{journal}{Phys. Rev. Lett.}}
  \textbf{\bibinfo{volume}{129}}, \bibinfo{pages}{186801}
  (\bibinfo{year}{2022}).

\bibitem{Zha22.2}
\bibinfo{author}{Zhang, C.-L.}, \bibinfo{author}{Liang, T.},
  \bibinfo{author}{Kaneko, Y.}, \bibinfo{author}{Nagaosa, N.} \&
  \bibinfo{author}{Tokura, Y.}
\newblock \bibinfo{title}{{Giant Berry curvature dipole density in a
  ferroelectric Weyl semimetal}}.
\newblock \emph{\bibinfo{journal}{npj Quantum Mater.}}
  \textbf{\bibinfo{volume}{7}}, \bibinfo{pages}{103} (\bibinfo{year}{2022}).

\bibitem{Min23.1}
\bibinfo{author}{Min, L.}, \bibinfo{author}{Tan, H.}, \bibinfo{author}{Xie,
  Z.}, \bibinfo{author}{Miao, L.}, \bibinfo{author}{Zhang, R.},
  \bibinfo{author}{Lee, S.~H.}, \bibinfo{author}{Gopalan, V.},
  \bibinfo{author}{Liu, C.-X.}, \bibinfo{author}{Alem, N.},
  \bibinfo{author}{Yan, B.} \& \bibinfo{author}{Mao, Z.}
\newblock \bibinfo{title}{{Strong room-temperature bulk nonlinear Hall effect
  in a spin-valley locked Dirac material}}.
\newblock \emph{\bibinfo{journal}{Nat. Commun.}} \textbf{\bibinfo{volume}{14}},
  \bibinfo{pages}{364} (\bibinfo{year}{2023}).

\bibitem{Du19.1}
\bibinfo{author}{Du, Z.~Z.}, \bibinfo{author}{Wang, C.~M.},
  \bibinfo{author}{Li, S.}, \bibinfo{author}{Lu, H.-Z.} \&
  \bibinfo{author}{Xie, X.~C.}
\newblock \bibinfo{title}{{Disorder-induced nonlinear Hall effect with
  time-reversal symmetry}}.
\newblock \emph{\bibinfo{journal}{{Nat.\ Commun.}}}
  \textbf{\bibinfo{volume}{10}}, \bibinfo{pages}{3047} (\bibinfo{year}{2019}).

\bibitem{Xia19.1}
\bibinfo{author}{Xiao, C.}, \bibinfo{author}{Zhou, H.} \& \bibinfo{author}{Niu,
  Q.}
\newblock \bibinfo{title}{{Scaling parameters in anomalous and nonlinear Hall
  effects depend on temperature}}.
\newblock \emph{\bibinfo{journal}{Phys. Rev. B}}
  \textbf{\bibinfo{volume}{100}}, \bibinfo{pages}{161403}
  (\bibinfo{year}{2019}).

\bibitem{Zha21.1}
\bibinfo{author}{Zhang, C.-L.}, \bibinfo{author}{Liang, T.},
  \bibinfo{author}{Bahramy, M.~S.}, \bibinfo{author}{Ogawa, N.},
  \bibinfo{author}{Kocsis, V.}, \bibinfo{author}{Ueda, K.},
  \bibinfo{author}{Kaneko, Y.}, \bibinfo{author}{Kriener, M.} \&
  \bibinfo{author}{Tokura, Y.}
\newblock \bibinfo{title}{{Berry curvature generation detected by Nernst
  responses in ferroelectric Weyl semimetal}}.
\newblock \emph{\bibinfo{journal}{{Proc. Natl. Acad. Sci. U.S.A.}}}
  \textbf{\bibinfo{volume}{118}}, \bibinfo{pages}{e2111855118}
  (\bibinfo{year}{2021}).

\bibitem{Sun15.1}
\bibinfo{author}{Sun, Y.}, \bibinfo{author}{Wu, S.-C.}, \bibinfo{author}{Ali,
  M.~N.}, \bibinfo{author}{Felser, C.} \& \bibinfo{author}{Yan, B.}
\newblock \bibinfo{title}{{Prediction of Weyl semimetal in orthorhombic
  MoTe$_2$}}.
\newblock \emph{\bibinfo{journal}{Phys. Rev. B}} \textbf{\bibinfo{volume}{92}},
  \bibinfo{pages}{161107} (\bibinfo{year}{2015}).

\bibitem{Qi16.1}
\bibinfo{author}{Qi, Y.}, \bibinfo{author}{Naumov, P.~G.},
  \bibinfo{author}{Ali, M.~N.}, \bibinfo{author}{Rajamathi, C.~R.},
  \bibinfo{author}{Schnelle, W.}, \bibinfo{author}{Barkalov, O.},
  \bibinfo{author}{Hanfland, M.}, \bibinfo{author}{Wu, S.-C.},
  \bibinfo{author}{Shekhar, C.}, \bibinfo{author}{Sun, Y.},
  \bibinfo{author}{S\"u{\ss}, V.}, \bibinfo{author}{Schmidt, M.},
  \bibinfo{author}{Schwarz, U.}, \bibinfo{author}{Pippel, E.},
  \bibinfo{author}{Werner, P.}, \bibinfo{author}{Hillebrand, R.},
  \bibinfo{author}{F\"orster, T.}, \bibinfo{author}{Kampert, E.},
  \bibinfo{author}{Parkin, S.}, \bibinfo{author}{Cava, R.~J.},
  \bibinfo{author}{Felser, C.}, \bibinfo{author}{Yan, B.} \&
  \bibinfo{author}{Medvedev, S.~A.}
\newblock \bibinfo{title}{{Superconductivity in Weyl semimetal candidate
  MoTe$_2$}}.
\newblock \emph{\bibinfo{journal}{{Nat.\ Commun.}}}
  \textbf{\bibinfo{volume}{7}}, \bibinfo{pages}{11038} (\bibinfo{year}{2016}).

\bibitem{Son16.1}
\bibinfo{author}{Song, Q.}, \bibinfo{author}{Pan, X.}, \bibinfo{author}{Wang,
  H.}, \bibinfo{author}{Zhang, K.}, \bibinfo{author}{Tan, Q.},
  \bibinfo{author}{Li, P.}, \bibinfo{author}{Wan, Y.}, \bibinfo{author}{Wang,
  Y.}, \bibinfo{author}{Xu, X.}, \bibinfo{author}{Lin, M.},
  \bibinfo{author}{Wan, X.}, \bibinfo{author}{Song, F.} \&
  \bibinfo{author}{Dai, L.}
\newblock \bibinfo{title}{{The in-plane anisotropy of WTe2 investigated by
  angle-dependent and polarized Raman spectroscopy}}.
\newblock \emph{\bibinfo{journal}{{Sci. Rep.}}} \textbf{\bibinfo{volume}{6}},
  \bibinfo{pages}{29254} (\bibinfo{year}{2016}).

\bibitem{Bro66.1}
\bibinfo{author}{Brown, B.~E.}
\newblock \bibinfo{title}{{The crystal structures of WTe$_2$ and
  high-temperature MoTe$_2$}}.
\newblock \emph{\bibinfo{journal}{{Acta Cryst.}}}
  \textbf{\bibinfo{volume}{20}}, \bibinfo{pages}{268--274}
  (\bibinfo{year}{1966}).

\bibitem{Zho18.1}
\bibinfo{author}{Zhong, S.}, \bibinfo{author}{Tiwari, A.},
  \bibinfo{author}{Nichols, G.}, \bibinfo{author}{Chen, F.},
  \bibinfo{author}{Luo, X.}, \bibinfo{author}{Sun, Y.} \&
  \bibinfo{author}{Tsen, A.~W.}
\newblock \bibinfo{title}{{Origin of magnetoresistance suppression in thin
  $\ensuremath{\gamma}\text{\ensuremath{-}}\mathrm{MoT}{\mathrm{e}}_{2}$}}.
\newblock \emph{\bibinfo{journal}{{Phys.\ Rev.\ B}}}
  \textbf{\bibinfo{volume}{97}}, \bibinfo{pages}{241409}
  (\bibinfo{year}{2018}).

\bibitem{Wu16.3}
\bibinfo{author}{Wu, Y.}, \bibinfo{author}{Mou, D.}, \bibinfo{author}{Jo,
  N.~H.}, \bibinfo{author}{Sun, K.}, \bibinfo{author}{Huang, L.},
  \bibinfo{author}{Bud'ko, S.~L.}, \bibinfo{author}{Canfield, P.~C.} \&
  \bibinfo{author}{Kaminski, A.}
\newblock \bibinfo{title}{{Observation of Fermi arcs in the type-II Weyl
  semimetal candidate ${\mathrm{WTe}}_{2}$}}.
\newblock \emph{\bibinfo{journal}{Phys. Rev. B}} \textbf{\bibinfo{volume}{94}},
  \bibinfo{pages}{121113} (\bibinfo{year}{2016}).

\bibitem{Li17.1}
\bibinfo{author}{Li, P.}, \bibinfo{author}{Wen, Y.}, \bibinfo{author}{He, X.},
  \bibinfo{author}{Zhang, Q.}, \bibinfo{author}{Xia, C.}, \bibinfo{author}{Yu,
  Z.-M.}, \bibinfo{author}{Yang, S.~A.}, \bibinfo{author}{Zhu, Z.},
  \bibinfo{author}{Alshareef, H.~N.} \& \bibinfo{author}{Zhang, X.-X.}
\newblock \bibinfo{title}{{Evidence for topological type-II Weyl semimetal
  WTe$_2$}}.
\newblock \emph{\bibinfo{journal}{Nat. Commun.}} \textbf{\bibinfo{volume}{8}},
  \bibinfo{pages}{2150} (\bibinfo{year}{2017}).

\bibitem{Hau17.2}
\bibinfo{author}{Haubold, E.}, \bibinfo{author}{Koepernik, K.},
  \bibinfo{author}{Efremov, D.}, \bibinfo{author}{Khim, S.},
  \bibinfo{author}{Fedorov, A.}, \bibinfo{author}{Kushnirenko, Y.},
  \bibinfo{author}{van~den Brink, J.}, \bibinfo{author}{Wurmehl, S.},
  \bibinfo{author}{B\"uchner, B.}, \bibinfo{author}{Kim, T.~K.},
  \bibinfo{author}{Hoesch, M.}, \bibinfo{author}{Sumida, K.},
  \bibinfo{author}{Taguchi, K.}, \bibinfo{author}{Yoshikawa, T.},
  \bibinfo{author}{Kimura, A.}, \bibinfo{author}{Okuda, T.} \&
  \bibinfo{author}{Borisenko, S.~V.}
\newblock \bibinfo{title}{{Experimental realization of type-II Weyl state in
  noncentrosymmetric ${\mathrm{TaIrTe}}_{4}$}}.
\newblock \emph{\bibinfo{journal}{Phys. Rev. B}} \textbf{\bibinfo{volume}{95}},
  \bibinfo{pages}{241108} (\bibinfo{year}{2017}).

\bibitem{Wu16.1}
\bibinfo{author}{Wu, L.}, \bibinfo{author}{Patankar, S.},
  \bibinfo{author}{Morimoto, T.}, \bibinfo{author}{Nair, N.~L.},
  \bibinfo{author}{Thewalt, E.}, \bibinfo{author}{Little, A.},
  \bibinfo{author}{Analytis, J.~G.}, \bibinfo{author}{Moore, J.~E.} \&
  \bibinfo{author}{Orenstein, J.}
\newblock \bibinfo{title}{{Giant anisotropic nonlinear optical response in
  transition metal monopnictide Weyl semimetals}}.
\newblock \emph{\bibinfo{journal}{{Nat.\ Phys.}}}
  \textbf{\bibinfo{volume}{13}}, \bibinfo{pages}{350} (\bibinfo{year}{2016}).

\bibitem{Wan22.2}
\bibinfo{author}{Wang, J.}, \bibinfo{author}{Wang, H.}, \bibinfo{author}{Chen,
  Q.}, \bibinfo{author}{Qi, L.}, \bibinfo{author}{Zheng, Z.},
  \bibinfo{author}{Huo, N.}, \bibinfo{author}{Gao, W.}, \bibinfo{author}{Wang,
  X.} \& \bibinfo{author}{Li, J.}
\newblock \bibinfo{title}{{A Weyl semimetal WTe2/GaAs 2D/3D Schottky diode with
  high rectification ratio and unique photocurrent behavior}}
  \textbf{\bibinfo{volume}{121}}, \bibinfo{pages}{103502}
  (\bibinfo{year}{2022}).

\end{thebibliography}

\end{document}